\documentclass[12pt]{article}
\usepackage{amssymb,amsmath,amscd}
\makeatletter

\@addtoreset{equation}{section}
\def\section{\@startsection {section}{1}{\z@}{-2.25ex plus -1ex minus
 -.2ex}{1.0ex plus .2ex}{\large\bf}}
\def\subsection{\@startsection{subsection}{2}{\z@}{-2.0ex plus%
 -1ex minus -.2ex}{0.5ex plus .2ex}{\bf}}

\def\Ad{\mathrm{Ad}}

\def\bmu{{\mbox{\boldmath $\mu$}}}

\def\bphi{{\mbox{\boldmath $\phi$}}}
\newcommand{\pp}{p}

\newcommand{\inv}[0]{{-1}}

\newcommand{\cif}[0]{\mathcal{C}^\infty}

\newcommand{\tenltimes}[0] {\mbox{$\subset\!\!\!\!\!\!\times$}}

\newcommand{\ee}[0]{\epsilon}

\newcommand{\surfb}[0]{S_{g,n}^\infty}
\newcommand{\xb}[0]{x_\infty}

\newcommand{\surf}[0]{S_{g,n}}

\def\LLor{\tilde L_3^\uparrow}
\def\PPoi{\tilde P_3^\uparrow }

\newcommand{\prgr}{G\ltimes \mathfrak{g}^*}

\def\ba{{\mbox{\boldmath $a$}}}
\def\bx{{\mbox{\boldmath $x$}}}
\def\bX{{\mbox{\boldmath $X$}}}
\def\bs{{\mbox{\boldmath $s$}}}

\def\bj{{\mbox{\boldmath $j$}}}
\def\bz{{\mbox{\boldmath $z$}}}
\def\pbj{{\mbox{\bf \j} }}
\def\bk{{\mbox{\boldmath $k$}}}
\def\bp{{\mbox{\boldmath $p$}}}

\def\bphi{\bar \varphi}

\newcommand{ \vecj} {\vec{\mbox{\j}}}

\newcommand{\gothc}{\mathfrak c }

\newcommand{\ZZ}{\mathbb{Z}}

\newcommand{\RR}{\mathbb{R}}

\newcommand{\MM}{\mathbb{M}}

\newcommand{\gothf}{\mathcal F }

\newtheorem{theorem}{Theorem}[section]
\newtheorem{lemma}[theorem]{Lemma}

\def\bea{\begin{eqnarray}}
\def\eea{\end{eqnarray}}
\def\bmz{\left(\begin{array}{2,2}}
\def\emz{\end{array}\right)}
\def\bmd{\left(\begin{array}{3,3}}
\def\emd{\end{array}\right)}

\newcommand{\mi}[0]{{M_i}}
\newcommand{\ai}[0]{{A_i}}
\newcommand{\bi}[0]{{B_i}}

\newcommand{\aj}[0]{{A_j}}
\newcommand{\bjj}[0]{{B_j}}

\newcommand{\me}[0]{{M_1}}

\newcommand{\mf}[0]{{M_n}}

\begin{document}
\parskip 6pt
\parindent 0pt
\begin{flushright}
EMPG-05-08\\
\end{flushright}

\begin{center}
\baselineskip 24 pt {\Large \bf  Boundary conditions and symplectic structure in the Chern-Simons formulation of (2+1)-dimensional gravity}

\baselineskip 16 pt

\vspace{.7cm} {{ C.~Meusburger}\footnote{\tt  cmeusburger@perimeterinstitute.ca}\\
Perimeter Institute for Theoretical Physics\\
31 Caroline Street North,
Waterloo, Ontario N2L 2Y5, Canada\\

\vspace{.5cm}
{ B.~J.~Schroers}\footnote{\tt bernd@ma.hw.ac.uk} \\
Department of Mathematics, Heriot-Watt University \\
Edinburgh EH14 4AS, United Kingdom } \\

\vspace{0.5cm}

{ May   2005}

\end{center}

\begin{abstract}

\noindent We propose a description of open universes in the
Chern-Simons formulation of (2+1)-dimensional gravity where
spatial infinity is implemented as a puncture. At this puncture,
additional variables are introduced which lie in  the cotangent
bundle of  the Poincar\'e group, and coupled minimally to the
Chern-Simons gauge field. We apply this description of spatial
infinity to open universes of general genus and with an arbitrary
number of massive spinning particles. Using results of \cite{we4}
we  give a finite dimensional description of the phase space and
determine its symplectic structure. In the special case of a genus
zero universe with spinless particles,  we compare our result to
the symplectic structure computed by Matschull in the metric
formulation of (2+1)-dimensional gravity. We comment on the
quantisation of the phase space and derive a quantisation
condition for the total mass and spin of an open universe.
\end{abstract}

\section{Introduction}

The goal of this paper is to introduce a mathematically
transparent treatment of spatial infinity for open universes in
the Chern-Simons formulation of (2+1)-dimensional gravity
\cite{AT,
  Witten1}  with vanishing cosmological constant and to
determine the phase space structure of the resulting theory. We
consider universes of general genus $g$ and with $n$ massive
spinning particles and work with the Hamiltonian formulation of
the Chern-Simons formalism.

In general terms, the  boundary requirement for an open universe
with vanishing cosmological constant which we study in this paper is
that  there be  an asymptotic region of  spacetime
which has the geometry of a spinning cone \cite{DJT}.  The implementation
of this requirement in any  Hamiltonian  formulation of gravity
requires the introduction of boundary degrees of freedom,
defined in the asymptotic  region of spacetime.
These boundary degrees of freedom  should then
be coupled to the bulk degrees of freedom in such a way that the
resulting equations of motion ensure the required conical geometry
in the asymptotic region.

The precise
formulation of the boundary requirement
 and its  implementation in the metric
formulation of gravity in (2+1) dimensions are  discussed in detail in
\cite{Matschull1}, which is a key reference for this paper.
Technically, the treatment  in \cite{Matschull1} is quite involved
since it is formulated in terms of conditions on the metric
in a neighbourhood of spatial infinity.
It requires the introduction of
infinitely many  boundary degrees of freedom, mathematically
represented by fields defined in a neighbourhood of spatial infinity.
However, after dividing out by gauge degrees of freedom the total
phase of the theory turns out to be finite dimensional.

In our  treatment of the boundary, the
number of degrees of freedom associated to the boundary is finite
from the start. This is appropriate for the physics of a
spatial boundary
in (2+1)-dimensional gravity with vanishing cosmological constant,
but should be contrasted with the situation for negative cosmological
constant, where one expects infinitely many degrees of freedom
to be associated to the boundary of a spacetime containing a black hole
 \cite{Carlipblack,BG}.
Working  in the formulation of  (2+1) dimensional gravity as
Chern-Simons theory  with the universal cover $\PPoi$ of the
Poincar\'e group  as  gauge group, we compactify spatial infinity
to a point and model it by a  puncture on the surface $S$. More
precisely, we  consider universes of topology $\RR \times \surfb$,
where $\surfb$ is a oriented  surface of genus $g$ with $n$
punctures representing massive, spinning particles. The
superscript $\infty$ refers to the  additional puncture which
represents the boundary at spatial infinity.

For the  first $n$
punctures representing particles we employ the standard treatment
of punctures in Chern-Simons theory \cite{Witten2}.
We introduce additional variables
which lie in coadjoint orbits of the $\PPoi$,
equipped  with their standard symplectic structure,
and couple them to the gauge field via minimal coupling. Each  coadjoint
orbit is labelled by the mass and spin of  the associated
particle.   It is explained in \cite{deSousaGebert} why
this treatment leads to
to the correct incorporation of  particles in the Chern-Simons
formulation of (2+1)-dimensional gravity. An immediate consequence of
the minimal coupling to the gauge field is that the
curvature of the Chern-Simons gauge field develops a delta-function
singularity  at  each puncture with a coefficient
which lies  in the  coadjoint orbit associated to the puncture.

For the puncture representing spatial infinity, in the following
referred to as the distinguished puncture, our treatment differs from
the standard approach.  We introduce variables which lie
 in the cotangent bundle of $\PPoi$, equipped  with its
canonical symplectic structure,
 and again couple minimally to the gauge field. Justifying
this mathematical description of spatial infinity is one of the
main tasks of this paper. In this introduction
we only point out that our model
leads to a curvature singularity  of the Chern-Simons gauge field  at
the distinguished
puncture with a Lie algebra valued coefficient. However, in contrast
to the ordinary punctures
the Lie algebra valued coefficient is not {\em a priori}
 restricted to a fixed adjoint orbit. 
Thus one may think of our distinguished puncture as describing
an auxiliary
particle whose mass and spin are  not fixed parameters but
additional variables. When  all the constraints are imposed,
the mass and spin of this auxiliary  particle turn out  to be
minus the total  mass and spin of the universe.

Having introduced the distinguished puncture and justified its
interpretation
as a model for spatial infinity, we parametrise the phase
space of the theory and determine its symplectic structure.
Here we rely heavily on the paper \cite{we4} where, using the
method of \cite{AMII},
the phase space structure of Chern-Simons theory with a distinguished
puncture   is analysed in detail for gauge
groups  of the form $\prgr$. We derive some physical
consequences  of our model and  relate our results to other approaches.
In particular, we discuss in detail the relation between our
description
of the phase space
and the results  derived in the metric formulation of
gravity in \cite{Matschull1}. In that paper, Matschull considers
$n$ spinless particles coupled to gravity.
He parametrises the phase space in terms of
variables
assigned to an oriented graph embedded into a  spatial slice of
spacetime.
We show how to reproduce his result
for a particular graph from our description of the phase space
in the special case of vanishing genus  $g=0$ and $n$ spinless particles.

We should also   comment
on the relation between this paper and our investigation of the
 phase space of open universes  in  the Chern-Simons formulation
of (2+1)-dimensional gravity  in \cite{we1}.
There we  considered universes of topology  $\RR\times (S_{g,
  n}\setminus D)$, where $S_{g , n}\setminus D$ is a surface of
genus $g$ with $n$ punctures and a disc removed.
We proposed a parametrisation  of the phase space  in terms of
finitely many holonomy variables  and
 computed the Poisson structure on this parameter space
 by applying a  technique developed by Fock
 and Rosly \cite{FR}. However, the derivation of the 
 Poisson structure given in
 \cite{we1} was not based on  a  field-theoretical treatment of 
boundary degrees of freedom and  relied  partly on symmetry arguments.
It turns out that both the symplectic structure associated
to the Poisson structure found in \cite{we1} and the symplectic
structure
we compute in this paper can be split into a piece that describes
 the centre of mass motion of the universe relative to an asymptotic observer
and a piece which describes the relative dynamics of the contents of
the universe.  While the piece  describing the relative dynamics is
the same in both symplectic structures, the symplectic structures for
the centre of mass motion differ. We  describe and comment on  this
difference  in Sect.~\ref{flower}.

The outline of the paper is as follows. In Sect.~2 we give a brief
review of the Chern-Simons formulation of (2+1)-dimensional
gravity and introduce our notation and  conventions. In Sect.~3 we
review  in some detail the metric of a spinning cone which
surrounds massive  point particles  with spin in (2+1)-dimensional
gravity and describe   how this geometry is captured in the
Chern-Simons formulation. This section provides important
background for Sect.~4, where
 we  review the boundary conditions at spatial infinity, and
explain how they lead to the description of spatial infinity as a
distinguished puncture. Having formulated our boundary condition
for spatial infinity, we then show how this condition can be
implemented via an action functional.
 In Sect.~5 we give a
finite-dimensional parametrisation of the phase space  and
determine the symplectic structure in terms of this
parametrisation. The mathematical results in this section mostly
follow directly from our paper \cite{we4}, but the physical
interpretation, particularly in the discussion of gauge fixing, is
new. In Sect.~6 we relate our description of the phase to that of
Matschull in the special case of vanishing genus  $g=0$ and $n$
spinless particles. Sect.~7 contains remarks about the
quantisation of our phase space. In particular,  we derive a
quantisation condition for the total mass and spin of the
universe. Sect.~8 contains our conclusions, and in the appendix we
give a brief summary of relevant results from \cite{Matschull1}
adapted to our conventions.


\section{The Chern-Simons formulation of (2+1)-dimensional gravity}

\subsection{Setting and conventions}

\label{setting}

We begin with a brief review of the Chern-Simons formulation of
(2+1)-dimensional gravity with vanishing cosmological constant for
closed universes, referring the reader to \cite{we1,Carlipbook}
for more details. Spacetime is assumed to have the topology $\RR
\times \surf$, where
 $S_{g,n}$ is  a two-dimensional, closed and
 oriented manifold of genus $g\geq 0 $ with   $n\geq 0$
punctures. We introduce a  coordinate $x^0$ on
$\RR$ and sometimes we use local coordinates
$x=(x^1,x^2)$ on
$\surf$. Differentiation with respect to $x^0, x^1$ and
$x^2$ is written as  $\partial_0,\partial_1$ and $\partial_2$.
The  coordinates of the $n$  punctures on $\surf$
are  $x_{(1)},\ldots,x_{(n)}$.
The symbol  $d$ stands for the
total  exterior derivative of any function or form. Sometimes we
consider functions or forms on   $\RR\times\surf$  and take
exterior derivatives with respect to the dependence on $\surf$ only;
such derivatives are then denoted
$d_S$.
Throughout the paper we  use units in which the
speed of light is $1$. In (2+1)-dimensional  gravity
Newton's constant has dimensions of inverse mass, allowing us to
measure masses in units of $(8\pi G)^{-1}$.

In the Chern-Simons formulation, (2+1)-dimensional gravity with
vanishing cosmological constant is written as a gauge theory with
the (2+1)-dimensional Poincar\'e group or one of its covers as
gauge group. More precisely, writing $L_3^\uparrow$ and
$P_3^\uparrow =L_3^\uparrow\ltimes\mathbb{R}^3$ respectively for
the proper orthochronous Lorentz and Poincar\'e group and
$\tilde{L}_3^\uparrow$, $\PPoi=\tilde{L}_3^\uparrow\ltimes
\mathbb{R}^3$ for their universal covers, we are going to take
$\PPoi$ as a gauge group in the following.  In order to define the
semi-direct product explicitly we need to specify how $\LLor$ acts
on $\RR^3$. We do this by identifying $\RR^3$ with the dual of the
Lie algebra $so(2,1)$ of  $\LLor$. The group $\PPoi$ is then of
the form $\prgr$ with $G=\LLor$ acting on the dual of its Lie
algebra in the coadjoint action. This is the formulation we are
going to adopt, allowing us to apply the results of the papers
\cite{we4,we2,we3} where various aspects of Chern-Simons theory
with gauge group $\prgr$ are studied in detail. Since the Lie
algebra $so(2,1)$ is semi-simple, however, we can simplify
notation by using  the Killing form $\eta$ on $so(2,1)$ to
identify it with its dual. Explicitly, we introduce
 generators $J_0,J_1,J_2$ of $so(2,1)$ normalised so that
\bea \eta(J_a,J_b)=\eta_{ab}=\text{diag}(1,-1,-1) \eea and
satisfying the commutation relations \bea [J_a,J_b]=\epsilon_{abc}
J^c, \eea where we use the convention $\epsilon_{012}=1$ and
raised indices with $\eta^{ab}=\eta_{ab}$. To obtain a set of
generators for  the Lie algebra $\text{Lie}\;\PPoi=iso(2,1)$ we
introduce additional generators $P_a$, $a=0,1,2$,  which commute
with each other and transform in the adjoint representation under
the generators $J_a$ $a=0,1,2$. The complete commutation relations
are then given by
 \bea
\label{poinccomm}
 [P_a,P_b] =0,
\quad[J_a,J_b]=\epsilon_{abc} J^c,\quad
[J_a,P_b]=\epsilon_{abc}P^c. \eea The elements $P^0,P^1,P^a$
generate translations in time and space while $J_0$ generates
rotations and $J_1$ and $J_2$ Lorentz boosts. The generators $J_0$
and $P^0$ of spatial rotations and time translations form an
abelian  Cartan subalgebra of $iso(2,1)$ which we denote by
$\gothc$ in the following. Note that, because of the Lorentzian
signature, \bea \label{signcon} [J_0,P_1]=-P_2 \quad \text{and}
\quad  [J_0,P_2]=P_1, \eea so that $J_0$ generates rotations in
the mathematically negative sense in the space spanned by the
translation generators $P_1$ and $P_2$.

In this paper  only  the Cartan subalgebra $\gothc$ plays a
role,
but we should point out that there are   non-conjugate Cartan
subalgebras in $iso(2,1)$, and that some elements in $iso(2,1)$
cannot be conjugated into any Cartan subalgebra. The general situation
 for the Lie algebras of groups  of the form $\prgr$ is summarised in
 \cite{we4}, where we also list references. Adopting the terminology
used for conjugacy classes of $SO(2,1)$, we call elements of
$iso(2,1)$ which are
conjugate to $\mu J_0 +s P_0$ for $\mu,s\in \RR, \mu \neq 0$, elliptic
elements and the Cartan subalgebras conjugate to $\gothc$ elliptic
Cartan subalgebras.
A second family of Cartan subalgebras, called hyperbolic Cartan
subalgebras
of $iso(2,1)$,  is obtained from
 the Cartan subalgebra spanned by $J_1$ and $P_1$ by conjugation.
Hyperbolic and elliptic Cartan subalgebras cannot be conjugated
into each other.  Finally, there are elements of the form
$h\cdot(\alpha(J_0\pm J_1) +\beta (P_0\pm P_1))\cdot h^\inv$, for
$\alpha,\beta \in \RR$, $h\in\PPoi$ which are not in any Cartan
subalgebra \cite{we4, Knapp}. We call these elements of $iso(2,1)$
parabolic.

For the definition of  the  Chern-Simons action
we require  a non-degenerate, invariant
  bilinear form on the Lie algebra. For the Lie algebra
 $iso(2,1)$ we have the pairing
\bea \label{inprod} \langle J_a, P^b\rangle = \delta_a^b, \quad
\langle J_a, J_b\rangle = \langle P^a,P^b\rangle = 0. \eea Note
that this pairing is canonically defined on the Lie algebra of
any Lie group of the form $\prgr$. It is worth stressing that the
earlier identification of $so(2,1)$ with $so(2,1)^*$ via $\eta$ is
only used to formulate the theory in a more familiar form; the
Chern-Simons formulation of gravity  requires the bilinear form
$\langle, \rangle$, but not the Killing form $\eta$.

With our conventions  and the parametrisation \bea
(u,\ba)\in\PPoi\qquad\text{with}\qquad u\in\tilde L_3^\uparrow,\;
\ba\in\mathbb{R}^3, \eea
 the group
multiplication law in $\PPoi$ takes the form \bea
\label{groupmult}
 (u_1,\ba_1)\cdot(u_2,\ba_2)=(u_1\cdot u_2,\ba_1+\Ad(u_1)\ba_2)
\eea with $\Ad(u)$ denoting the $L_3^\uparrow$ element associated
to $u\in \tilde L_3^\uparrow$. The $\Ad(u)$ action on $\ba$ is the
familiar representation of an $\LLor$ element  as an $SO(2,1)$ matrix.
In this paper we will often need to refer to the abelian subgroup of
$\PPoi$ generated by the Cartan subalgebra $\gothc$; we will
denote it by $T_\gothc$. Its elements are rotations and time
translations
 and can be parametrised as $(e^{-\Phi J_0}, T P_0)$ with $\Phi,
T \in \RR$, where the  minus sign is inserted to take into account
the remark after \eqref{signcon}.

\subsection{The Chern-Simons action}
Einstein's theory of gravity is a field theory for a metric $g$ on
the spacetime manifold $M$.  The starting point for the
 Chern-Simons formulation of (2+1)-dimensional
gravity  is  Cartan's point of view, where the theory is
formulated in terms of the (non-degenerate) dreibein of one-forms
$e_a$, $a=0,1,2$, and the spin connection one-forms $\omega_a$,
$a=0,1,2$. The dreibein is related to the metric via \bea
\label{metdreib} \eta^{ab} e_a\otimes e_b \, = g, \eea and the
one-forms $\omega_a$  are the coefficients  of the
spin connection in the expansion
 \bea
\omega = \omega^a J_a.
\eea In the
formulation of (2+1)-dimensional gravity as a Chern-Simons gauge
theory, dreibein and spin connection are combined into the
 Cartan connection \cite{Sharpe} or
Chern-Simons gauge field. This  is  a  one-form
with values in the Lie algebra $iso(2,1)$
\bea
\label{Cartan} A =
\omega^a J_a + e_a P^a,
\eea
whose  curvature
 \bea
\label{decomp} F = T_a  P^a + F_\omega^a\,J_a \eea combines the
curvature and the torsion of the spin connection:
\begin{align}
\label{spincurv} &F_\omega= \omega + \frac{1}{2}[\omega,\omega]=
F^a_\omega J_a, & &\;F_\omega^a = d\omega^a  + \frac{1}{2}
\epsilon^a_{\;bc} \omega^b\wedge \omega^c\\
\label{notorsion} &T_a = de_a+  \epsilon_{abc} \omega^b
e^c.\end{align}

The equations of motion are obtained by variation of an action
functional, the Chern-Simons action, which in the absence of
punctures (i.e. $n=0$) is given by \bea \label{CSaction} S_{CS}[A]
=\frac{1}{2} \int_M \langle A\wedge  d A\rangle
 +\frac{2}{3}\langle A \wedge A \wedge A\rangle.
\eea It is invariant under Chern-Simons gauge transformations
\begin{align}
\label{CSgt} A\mapsto\gamma A \gamma^\inv+\gamma d\gamma^\inv
\end{align}
for general functions $\gamma:M\rightarrow\PPoi$, which take the
place of diffeomorphisms in Einstein's formulation of the theory
\cite{Witten1}. Variation of  the Chern-Simons action
\eqref{CSaction} with respect to the gauge field $A$ yields the
condition \bea \label{flat} F=0. \eea Thus, in the absence of
matter both the spin connection $F_\omega$ and the torsion $T$
vanish. In 2+1 dimensions the vanishing of the spin connection is
equivalent to the vanishing of the Einstein tensor. Equation
\eqref{flat} is therefore equivalent to the Einstein equations.
However, due to the different role of degenerate dreibeins, the
precise relationship between Einstein's theory and the
Chern-Simons formulation is problematic. In the Cartan formulation
of Einstein's theory in 2+1 dimensions, the dreibein $e_a$ is
required to define a metric and therefore has to be
non-degenerate. The Chern-Simons formulation of (2+1)-dimensional
gravity uses the $\PPoi$-connection $A$ on $M$ as the basic field
and does not impose the  non-degeneracy of the dreibein. It is
argued in \cite{Matschull2}, see also \cite{SchS}
for an earlier work in the (1+1)-dimensional context,
 that this leads to global
differences in the structure of the phase spaces of the two
theories. The results in this paper are based on the Chern-Simons
formulation of (2+1)-dimensional gravity, although we will discuss
the link with the metric formulations extensively in
Sect.~\ref{metric}.

\section{Including particles}
\label{particle}

\subsection{Point particles and the metric of the spinning cone}

In order to motivate the coupling of particles to the Chern-Simons
gauge field $A$ we need to discuss  the metric which solves the
Einstein equations for the energy momentum tensor for a single
massive
 particle   with spin. We do this in some detail
here since this discussion will provide important background for
the formulation of  boundary conditions in our treatment of
open universes in the next section.

The metric which solves Einstein's equations  in the presence of a
single point particle of mass $\mu$ and spin $s$ was derived by
Deser, Jackiw and  t'Hooft \cite{DJT}. It is defined on $\RR^3$
with the $x^0$-axis removed, denoted $\RR^3\setminus \RR$ in the
following.
 In terms of polar coordinates $r,\varphi$ for $x\in \RR^2$
so that $\varphi$
has  the range  $[0,2\pi)$, and a third coordinate $t$  
the line element can be written
\begin{align}
\label{partmet} ds^2=(dt+\tfrac{s}{2\pi} d\varphi)^2-\frac{1}{(1-\tfrac{\mu}{2\pi})^2}dr^2-r^2d\varphi^2.
\end{align}
In this solution it is assumed that $0\leq \mu<2\pi$.
The metric \eqref{partmet} is flat everywhere except on the $x^0$  axis.
 The quickest way to understand  the geometry defined by this metric
is to introduce ``flat coordinates'' $T,R,\Phi$ via
\bea
\label{flatcoords}
\Phi=(1-\frac{\mu}{2\pi})\varphi, \quad T=t+\frac{s} {2\pi}\varphi, \quad
 R=\frac{r}{1-\frac{\mu}{2\pi}},
\eea in terms of which the line element takes the form of the
Minkowski metric \bea \label{flatline} ds^2=dT^2-dR^2-R^2d\Phi^2.
\eea However, the identification $\varphi\sim \varphi + 2 \pi$
implies that one  has to identify \bea \label{identification}
\Phi\sim \Phi +(2\pi -\mu),\quad T\sim T+s. \eea If we set $s=0$
we see that spatial surfaces $T=$const. are cones with deficit
angle $\mu$ and apex on the line defined by $R=0$.
 In the general case $s\neq 0$ the geometry defined by \eqref{flatline}
and \eqref{identification} is called a ``spinning cone'' with deficit
angle $\mu$ and time-offset $s$.

There is a systematic way of  introducing the coordinates
$T,R,\Phi$ discussed under the heading ``developing map''
in the
mathematical literature \cite{Thurston,Sharpe},
 but also explained in
\cite{Matschull1} for the context of (2+1)-dimensional gravity. We
start with the dreibein and spin connection corresponding to the
metric \eqref{partmet} given in \cite{deSousaGebert}
\begin{align}
\label{dreib2}
&e^0=dt+\tfrac{s}{2\pi}\,d\varphi & &\omega^0=\tfrac{\mu}{2\pi}d\varphi\\
&e^1=\tfrac{1}{1-\tfrac{\mu}{2\pi}}\cos\varphi \,dr-r\sin\varphi\,
d\varphi & &\omega^1=0\nonumber\\
&e^2=\tfrac{1}{1-\tfrac{\mu}{2\pi}}\sin\varphi \,dr+r\cos\varphi \, d\varphi & &\omega^2=0\qquad\qquad \mu\in[0,2\pi),s\in\RR\nonumber.
\end{align}
Away from the line $R=0$ the  corresponding $\PPoi$-gauge field
\bea \label{conegauge}
 A_p=e_a P^a+\omega^a J_a
\eea is flat and can therefore be trivialised locally. Since
$\RR^3\setminus \RR$ is not simply-connected it is not  possible
to trivialise $A_p$ everywhere in terms of a single
 $\PPoi$-valued function. However, defining the multi-valued function
\bea \label{gammafield2} \Gamma_p=
\left(e^{-\frac{\mu\varphi}{2\pi}J_0},
-(t+\tfrac{s\varphi}{2\pi})P_0-\tfrac{r}{1-\tfrac{\mu}{2\pi}}
\cos\varphi P_1-\tfrac{r}{1-\tfrac{\mu}{2\pi}}\sin\varphi
P_2\right) \eea we  have \bea \label{trivial} A_p=\Gamma_p d
\Gamma_p^\inv. \eea The translational part of the trivialising
function
defines an embedding  of $\RR^3\setminus\RR$ into  Minkowski space
$\MM^3$ such that the induced metric is \eqref{partmet}. Its
Lorentz component defines a Lorentz frame at every point of the
embedding. Explicitly, parametrising $\Gamma_p^\inv=(V,\bX)$ with
$V:\RR^3\setminus \RR \rightarrow\LLor$, $\bX:\RR^3\setminus \RR
\rightarrow \MM^3$
 we have
\bea \label{developing} V= e^{\frac{\mu\varphi}{2\pi}J_0}\qquad
\bX= T P_0 + R\cos\Phi P_1 + R\sin \Phi P_2 \eea in terms of the
coordinates $T,R,\Phi$ defined in \eqref{flatcoords}. It then
follows from \eqref{trivial} that $e^aP_a=\Ad(V^\inv)d\bX$ and
hence \bea ds^2=\eta_{ab}e^a e^b=\eta_{ab}dX^a dX^b \eea in
agreement with \eqref{flatline}. The embedding map
$ X:\RR^3\setminus\RR \rightarrow \MM^3$
 shows that one can construct the spinning cone
from Minkowski space $\MM^3$  by cutting out the  wedge  $2\pi
-\mu <\Phi < 2\pi$ and identifying  the half-planes
$\Phi=2\pi-\mu$ and $\Phi=0$  after shifting one by $s$ in the
$T$-direction. Equivalently, the spinning cone can be thought of
as a quotient of the universal cover of $\MM^3\setminus\RR$ by the
equivalence relation \eqref{identification}.

The function $\Gamma_p$ in \eqref{trivial}  is unique only up to
right-multiplication with a constant Poincar\'e element
$w^\inv\in\PPoi$ or, equivalently, left-multiplication of
$\Gamma_p^\inv$ with $w$: \bea \label{linvariance} \Gamma_p^\inv
\mapsto w \Gamma_p^\inv, \eea which leaves the gauge field
\eqref{trivial} invariant. Such a transformation corresponds to a
Poincar\'e transformation in the embedding
 Minkowski space. It leaves the dreibein and spin connection,
and hence the  metric invariant, but changes the way the cone is
embedded. In particular, transformations with elements
$w=(e^{-\Phi_0 J_0}, T_0 P_0)$ in the abelian  subgroup $T_\gothc$
 map $\Phi \mapsto \Phi +\Phi_0$
and $T\mapsto T+ T_0$.
Such transformations  rotate and translate the embedded cone into itself,
 but more general transformations $w\in \PPoi$ translate and boost the
cone's axis in the  Minkowski space. Note that in order to respect
the identification  \eqref{identification} the $\PPoi$-element $w$
itself should be viewed as an element of the quotient of $\PPoi$
by the relation \bea \label{phirange}
 w\sim w(e^{-(2\pi-\mu)J_0},sP_0).
\eea

\subsection{The spinning cone in the gauge theoretical formulation}

In  $\PPoi$-Chern-Simons theory the gauge field  $A_p$
\eqref{conegauge} should be identified  with any field obtained
from it via a gauge transformation \eqref{CSgt}. In particular,
$A_p$  is gauge equivalent to the singular gauge field \bea
\label{singpartgauge} \hat A_p = (\tfrac{\mu}{2\pi} J_0
+\tfrac{s}{2\pi} P_0) d\varphi \eea to which it is related via the
gauge transformation \bea \label{partgauge} A_p=\gamma_p\hat
A_p\gamma_p^\inv +\gamma_p d\gamma_p^\inv \eea with the
(single-valued) function \bea \label{regtriv}\gamma_p=(1, -t
P_0-\tfrac{r}{1-\tfrac{\mu}{2\pi}} \cos\varphi
P_1-\tfrac{r}{1-\tfrac{\mu}{2\pi}}\sin\varphi P_2). \eea  This
allows us to decompose the trivialising function
$\Gamma_p=\gamma_p\hat\Gamma_p$ into a single-valued part
\eqref{regtriv} and a multi-valued part \bea \hat \Gamma_p
=(e^{-\frac{\mu\varphi}{2\pi}J_0}, -\tfrac{s\varphi}{2\pi}P_0)
\eea which trivialises the singular gauge field
\eqref{singpartgauge} \bea \label{partgauge2} \hat A_p=\hat
\Gamma_p d\hat \Gamma_p^\inv. \eea  As the translational part of
$\hat \Gamma_p^\inv$ is simply $\tfrac{s\varphi}{2\pi}P_0$, the
function $\hat\Gamma_p^\inv$ embeds all points on the cone into
the singularity $R=0$. Unlike \eqref{developing}, this map is
degenerate and does not induce a metric on $\RR^3\setminus \RR$.
The singular gauge illustrates that the metric interpretation of a
connection $A_p$ is gauge dependent in $\PPoi$-gauge theory.

The natural gauge-covariant object associated  to a connection is
its curvature. For the singular gauge field $\hat A_p$ we find
\bea \label{parteq} \hat F_p(x ) =\left(\mu J_0 + s P_0\right)
dx^1\wedge dx^2 \delta^{(2)}(x). \eea Under gauge transformations
\eqref{CSgt} the curvature is conjugated, but since \bea
\gamma_p(r=0)=(1,-tP_0) \eea the formula \eqref{parteq} also gives
the curvature of $A_p$. Under a general gauge transformation
$\gamma$, the curvature $\hat F_p$ at the singularity changes to
\bea \label{gencurv} F_p = h \left(\mu J_0 + s P_0\right)  h^\inv
dx^1\wedge dx^2 \delta^{(2)}(x) \qquad  h(x^0)=\gamma(x^0,0).\eea
This formula shows that the gauge invariant object associated to a
particle with mass $\mu$ and spin $s$ is the coadjoint orbit of
the element $\mu J_0 + s P_0$ in the Cartan subalgebra $\gothc$,
identified with a corresponding adjoint orbit via the
non-degenerate form $\langle, \rangle$. Defining \bea
\label{coadorbit} T= k_a P^a + p^a J_a=h \left(\mu J_0 + s
P_0\right)  h^\inv \eea and parametrising $h=(v,\bx)$ we have the
explicit formulae \bea \label{porbit} p^a J_a =\mu \Ad(v)
J_0\qquad k_a P^a=[\bx,p^a J_a] +s\Ad(v)  P_0, \eea or, in terms
of the component vectors $\vec k =(k_0,k_1,k_2)$, $\vec x=
(x_0,x_1,x_2)$, $\vec p =(p_0,p_1,p_2)$ and $\hat p=\vec p/\mu$
\bea \vec k=\vec x \wedge \vec p + s\hat p. \eea The vectors $\vec
p$ and $\vec k$ have the physical interpretation of the momentum
and (generalised) angular momentum of a free relativistic
particle, see \cite{deSousaGebert} and \cite{we1} for a more
detailed discussion. From \eqref{porbit} it follows that they
satisfy  two relations, the mass and spin constraint
 \bea \label{partconstraint} p_ap^a=\mu^2 \qquad
p_ak^a=\mu s. \eea

We have seen that particles lead to curvature singularities in the
formulation of (2+1)-dimensional gravity as a $\PPoi$-gauge
theory. The physical information about the particle, its
three-momentum  $\vec p$ and angular momentum $\vec k$, are
encoded in the $iso(2,1)$-valued coefficient of the curvature at
the singularity. An alternative way of capturing the physical
information about the particle is the holonomy along a path
surrounding the particle. Such holonomies will play an important
role in this paper, and we
 therefore review them briefly.
The holonomy for an infinitesimal circle surrounding the puncture
in the singular gauge \eqref{singpartgauge} is \bea
\label{element0} \widehat{\text{Hol}}_p=(e^{-\mu J_0},-sP_0), \eea
and in the  general gauge \eqref{partgauge}  \bea \label{element}
\text{Hol}_p=h(e^{-\mu J_0},-sP_0)h^\inv. \eea Parametrising the
holonomy as \bea \label{holpara} \text{Hol}_p=(u,-\Ad(u)\bj) \eea
and setting $h=(v,\bx)$, we have \bea \label{ujpar}u=e^{-p^a
J_a}\quad \bj =(1-\Ad(u^\inv))\bx + s \Ad(v) P_0\eea  with $\bp$
defined as in \eqref{porbit}. The associated component vector
$\vecj=(j_0,j_1,j_2)$ is therefore given by \bea \vecj =
(1-\Ad(u^\inv))\vec x + s \hat p, \eea and the vectors $\vec
p,\vecj$ again satisfy the mass and spin constraint
 \bea \label{partconstraints} p_ap^a=\mu^2 \qquad
p_a j^a=\mu s. \eea  The mathematical motivation of the
parametrisation \eqref{holpara} comes from the Poisson-Lie
structure of $\PPoi$ and is explained in \cite{we1} and
\cite{we4}. Physically, $u$ can be viewed as a group valued
momentum and $\bj$ as a generalised angular momentum of a point
particle coupled to (2+1)-dimensional gravity. The relation
between $u$ and $\bj$ on the one hand and the free particle
momentum  $\bp$ and angular momentum $\bk$  is discussed in some
detail in \cite{we1}.

\subsection{Coupling particles to the Chern-Simons action}

There is a standard prescription for coupling coadjoint orbits to
the Chern-Simons action \eqref{CSaction}, valid for any gauge
group \cite{Witten2}. The coupling of particles to
(2+1)-dimensional gravity in the Chern-Simons formulation given in
\cite{deSousaGebert} follows this prescription. The kinetic term
for each particle is derived from the symplectic structure
canonically associated to the orbit \eqref{coadorbit}, and the
orbit parameter $h$ is coupled to the gauge field via minimal
coupling. We now consider a spacetime $M\approx\RR\times\surf$
with $n$ particles of mass and spin $\mu_i$ and $s_i$,
$i=1,\ldots,n$ and parametrise the associated coadjoint orbit as
\bea \label{partel} T_i=h_i(\mu_i J_0+ s_i P_0 ) h_i^\inv,\qquad
h_i\in\PPoi. \eea

The product structure $M\approx \RR\times\surf$ allows one to give
a Hamiltonian formulation of Chern-Simons theory coupled to
coadjoint orbits. For this, one decomposes the gauge field with
respect to the coordinate $x^0$  as
\begin{align}
\label{gaugedec}
A=A_0 dx^0+A_S,
\end{align}
where $A_S$ is an $x^0$-dependent and Lie algebra valued one-form
on the spatial surface $\surf$ and $A_0$ is a Lie algebra valued
function on $\RR\times\surf$. Correspondingly,  the curvature $F$
is given as the sum \bea \label{curvdec} F=dx^0\wedge(\partial_0
A_S - d_S A_0 + [A_0,A_S]) + F_S, \eea  of a term proportional to
$dx^0$ and the curvature two-form $F_S$ on $\surf$ \bea
\label{curvv} F_S=d_S A_S + A_S\wedge A_S. \eea In terms of this
decomposition, the action for the Chern-Simons formulation of
(2+1)-dimensional gravity coupled to particles can be written as
\bea \label{ordaction}
 S[A_S,A_0, h_i]&=&
\int_\RR dx^0\int_{\surf}
\tfrac{1}{2}\langle\partial_0 A_S\wedge A_S\rangle
 -\int_\RR dx^0 \sum_{i=1}^n \langle
\mu_i J_0+ s_i P_0,\, h_i^\inv\partial_0 h_i\rangle\\
& +&\int_\RR dx^0\int_{\surf}\langle A_0\,,\, F_S - \sum_{i=1}^n
T_i\delta^{(2)}(x-x_{(i)})dx^1\wedge dx^2\rangle . \nonumber \eea
The Lie algebra valued function $A_0$ plays the role of a Lagrange
multiplier and varying it leads to the constraint \bea
\label{curvconstraint} F_S(x^0,x)=\sum_{i=1}^n
T_i(x^0)\delta^{(2)}(x-x_{(i)})dx^1\wedge dx^2. \eea The evolution
equations obtained by varying $A_S$  are \bea
 \partial_0 A_S=d_SA_0+[A_S,A_0],\label{asvary}
\eea
and variation of $h_i$ implies
\bea
\partial_0 T_i=[T_i, A_0(x^0,x_{(i)})].\label{hivary}
\eea

\section{Spatial infinity as a distinguished puncture}

\subsection{Physical motivation}
\label{physmot} In this section we propose a Chern-Simons model
for open universes. Our approach is motivated by Matschull's
treatment of the boundary in
 the metric formulation of (2+1)-dimensional gravity  in
\cite{Matschull1}. That paper focuses on spacetime  manifolds of
topology $\RR \times N$, where
$N$   is $\RR^2$  with $n$ discs
removed. The asymptotic region  of $N$ represents spatial
infinity, and the boundary condition is the requirement that there
exists a neighbourhood of infinity
 $P_\infty \subset N$
 such that   the dreibein and spin connection
restricted to $P_\infty$ are those of a  spinning cone
\eqref{dreib2}  in suitable coordinates. This condition can be
viewed as the (2+1)-dimensional analogue of requiring that a
spacetime is asymptotically Minkowski space  in (3+1) dimensions.
In physical terms, it states that to an asymptotic observer, the
universe appears like a single particle of mass $\mu$ and spin
$s$. However,  the cone's deficit angle  $\mu$ and time offset $s$
are now promoted from constants to phase space variables. As
explained in \cite{Matschull1} and \cite{giukiezeh}
 the introduction of $\mu$ and $s$  as phase space variables
has to be accompanied by the introduction of  further variables in
order to obtain a symplectic phase space. These extra variables
specify the way the cone is embedded in an  Minkowski space. Equivalently,
 they specify the relation between a distinguished
reference frame associated to spatial infinity and the reference
frame of  an asymptotic observer.

Before explaining the preceding statement in detail,
  it is useful to consider
the analogous  situation in (3+1) dimensions, discussed in
in \cite{regte}.  There,  one usually imposes as a boundary
condition  that the metric asymptotically takes the form of the
(3+1)-dimensional Minkowski metric. The variables associated to
spatial infinity are a momentum four-vector and a position
four-vector. The components of these two vectors are conjugate to
each other and specify, respectively, the universe's total energy
and momentum and a time and position with respect to a
distinguished reference Minkowski frame associated to spatial
infinity. In other words, they specify a Poincar\'e transformation
relating the distinguished Minkowski frame defined by the
asymptotic Minkowski metric to the Minkowski frame of an
asymptotic observer. Concretely, one can think of such a
distinguished Minkowski frame as being given by a set of very
distant fixed stars with respect to which an observer can specify
 time and  positions.

In (2+1) dimensions with the  boundary condition that the metric
is asymptotically conical, the geometry of  the asymptotic cone  selects
a family of frames whose time axis coincides with the cone's
  symmetry axis.
To distinguish one of these frames we need to fix an origin of time
and a direction in space by referring  to some physical event, in
analogy to the  (3+1)-dimensional  situation reviewed above.
 The variables associated to spatial
infinity then  specify a (2+1)-dimensional Poincar\'e
transformation with respect to  the distinguished reference
frame.   In the Chern-Simons
formulation, this distinguished reference  frame is given
by the    trivialising function $\Gamma_p$ in \eqref{trivial}.
 The  map  \eqref{developing} uses coordinates $T,R,\Phi$ which
implicitly refer to a standard reference frame of $\MM^3$ with the
point $T=R=0$ as the origin, with the time axis along the $T$-axis
and the $X^1$-axis in the direction $\Phi=0$. The $T$-axis is
uniquely associated to the embedded cone because it is its
symmetry axis; the point $T=0$ on that axis  and the orthogonal
direction $\Phi$ are fixed by mere convention. We can go from the
standard frame to another frame with the same $T$-axis by
rotations and time translations \bea \label{comamb} \Phi \mapsto
\Phi +\Phi_0, \qquad T\mapsto T+ T_0, \eea which correspond to
right-multiplication of the trivialising function \eqref{trivial}
by elements of the abelian subgroup $T_\gothc$.

More generally,
composing the embedding  with the Poincar\'e transformation $w$  as
in \eqref{linvariance} boosts and translates
the axis of the embedded cone with respect to the $T$-axis.
For non-trivial boosts, the axis of the embedded cone is no longer
parallel to the $T$-axis and therefore
 the intersection point of the axis with surfaces of
constant $T$ varies with $T$. The spacetime asymptotically appears like a
particle of mass $\mu$ and spin $s$ that moves with respect to an
 observer at rest  in the
reference Minkowski frame with coordinates $T,R,\Phi$. The
frames where the axis of the cone coincide with the $T$-axis can
therefore be viewed as centre of mass frames of the universe: the
spacetime asymptotically appears like  particle of mass $\mu$ and
spin $s$ at rest at the origin.
 The set of all
centre of mass frames is parametrised by a time shift $T_0$ and
an angle $\Phi_0$. These are the variables which will play the
role of
 conjugate variables to $\mu$ and $s$ as  in \cite{Matschull1}.
At this stage the variables $T_0$ and $\Phi_0$ merely parametrise
the set  of centre of mass frames relative to an arbitrarily
chosen standard frame. The standard frame only acquires a physical
significance and becomes distinguished
after gauge fixing in Sect. \ref{gaugefix}.

\subsection{The boundary condition at spatial infinity}

In this section we show how to implement this treatment of  the
boundary in the Chern-Simons formulation of (2+1)-dimensional
gravity, considering the general case of a spacetime of genus $g$
with $n$ massive, spinning particles. We shall argue shortly that
the boundary can be modelled by an additional puncture, which is
treated differently from the ordinary punctures representing
particles. In order to relate our treatment to the that using the
metric formalism in \cite{Matschull1}
 we begin by
by considering a universe of topology $\RR \times (\surf\setminus
D)$.

In the Chern-Simons formulation, the  equations of motion derived
from the action \eqref{ordaction} are evolution equations for the
connection $A_S$ and the orbit variables $h_i$. Boundary
conditions can be be imposed on either $A_S$ or its derivative
with respect to $x_0$. We seek boundary conditions on $A_S$ which,
together with the equations of motion,
 ensure that, in a suitable gauge,  solutions $A$ of the equations
of motion have the dreibein and spin connection of a spinning cone
in some neighbourhood $\RR\times P_\infty$ of the boundary of
$\RR\times(\surf\setminus D)$ which does not contain any
punctures. To make this statement precise, we introduce polar
coordinates $r,\bphi$ in $P_\infty$ as follows. Let  $\bphi$ be an
angular coordinate such that closed paths  parametrised by
increasing $\bphi \in [0,2\pi)$ encircle the removed disc $D$
clockwise, i.e. in the mathematically negative sense. Such paths
therefore encircle the ``rest of the universe''
 (handles and ordinary
punctures) in a mathematically positive sense, see
Fig~\ref{pi1gen}. The coordinate  $r$  is a function on $P_\infty$
which  increases monotonically  towards the boundary and takes
some large value at the boundary.  Then the requirement that a
solution of \eqref{flat} has the asymptotic form of a spinning
cone means that there exists a gauge such that the gauge field
takes the form \bea A_\infty = e_aP^a + \omega^a J_a \eea on
$\RR\times P_\infty$, with the dreibein and spin connection given
by
\begin{align}
\label{dreib22}
&e^0=dt+\tfrac{s}{2\pi}\,d\bphi & &\omega^0=\tfrac{\mu}{2\pi}d\bphi\\
&e^1=\tfrac{1}{1-\tfrac{\mu}{2\pi}}\cos\bphi \,dr-r\sin\bphi\,
d\bphi & &\omega^1=0\nonumber\\
&e^2=\tfrac{1}{1-\tfrac{\mu}{2\pi}}\sin\bphi \,dr+r\cos\bphi \, d\bphi &
&\omega^2=0,\nonumber
\end{align}
where $t$ is a  function of $x^0$.

Our boundary  condition is to require the existence of
a gauge such that the component $A_S$ of the connection $A$
  restricted to $P_\infty$  agrees with the restriction of
$A_\infty$  to $P_\infty$, i.e.
\bea \label{boundary} A_S=A_{\infty,S}  \;\text{ on} \quad
P_\infty. \eea
As  in the discussion  preceding
\eqref{singpartgauge} we can gauge transform this potential into
the simpler, but singular form
\bea
\label{singgaugeinf}
\hat A_{S}=(\tfrac{\mu}{2\pi} J_0+\tfrac{s}{2\pi} P_0)d\bphi.
\eea
The singular form of the gauge potential \eqref{singgaugeinf}
captures all  the gauge invariant information about the connection in
the region $P_\infty$. We can encode  this in a particularly simple fashion by
shrinking the disc $D$ to a point $x_\infty$, called the distinguished
puncture in the following,  and requiring the curvature to have a
delta-function singularity there. For this purpose
introduce polar coordinates $\rho,\varphi$ near the puncture  with
  $\varphi=2\pi-\bphi$ so that the path parametrised
by increasing $\varphi$ encircles the disc $D$ in a mathematically
positive sense and  $\rho$ is such that the worldline of the puncture
 is given by $\rho=0$.
 With
\bea
\label{singgaugeinff}
\hat A_S=-(\tfrac{\mu}{2\pi} J_0+\tfrac{s}{2\pi} P_0)d\varphi
\eea
the curvature now satisfies
\bea
\hat F_S=-(\mu J_0+ s P_0) dx^1 \wedge dx^2 \delta^{(2)}(x-\xb)
\eea
in the region $P_\infty$.  Under gauge transformations $\gamma_\infty:
P_\infty \rightarrow \PPoi$
in the region $P_\infty$, the singular
gauge potential $\hat A_{\infty,S}$ changes to
\bea
\label{boundgf3} A_{S,\infty}=\gamma_\infty\hat A_{S,\infty}
 \gamma_\infty^\inv +
\gamma_\infty d_S\gamma_\infty^\inv
\eea
and the curvature $\hat F_S$ gets conjugated by
 $h=\gamma_\infty(x_\infty)$. We therefore conclude that the
requirement that space has the asymptotic geometry of
a spinning cone can be implemented  in gauge-theoretical formulation
of (2+1)-dimensional gravity by adding a distinguished puncture
at $\xb$
representing spatial infinity and demanding that the
spatial part of the curvature
is
\bea
\label{singcoeff}
 F_S= -h(\mu J_0 +s P_0)h^\inv dx^1 \wedge dx^2 \delta^{(2)}(x-\xb)
\eea
for some $h\in \PPoi$,
 $\mu\in (0,2\pi) $ and  $s\in \RR$ and $h\in \PPoi$.

 The
 parameters $\mu,s$ are not fixed constants. Instead, they
 play the role of phase
space variables in our model.
We will drop the  the restriction
on the range of $\mu$ in the following but
shall see in the next subsection that   for solutions of the
equations of motion $\mu$ and $s$ are constant. We can therefore recover
solutions with the geometrical interpretation of a spinning cone
by restricting attention to solutions with
$\mu \in (0,2\pi)$.
More generally, however, we only require that the coefficient of the
curvature singularity in \eqref{singcoeff}
  is an elliptic element $T\in iso(2,1)$. According to our remarks in
Sect.~\ref{setting}, these can always be parametrised as
\begin{align}
\label{boundel} T=-h(\mu J_0+s P_0)h^\inv,
\end{align}
 for some $h \in \PPoi$ and $\mu,s \in \RR$, $\mu \neq 0$.

As anticipated at the beginning of the section, the set of all $T$
of the form \eqref{boundel} does not carry a canonical symplectic
structure when $\mu$ and $s$ are not fixed. We therefore need
additional variables associated to spatial infinity which ensure
that we obtain a symplectic structure. These variables should
 parametrise the embedding of the asymptotic cone in
Minkowski space. For this, we note that if a gauge field
$A=\Gamma_\infty d\Gamma_\infty^\inv$ is trivialised by a
(multi-valued) function $\Gamma_\infty:\RR\times
P_\infty\rightarrow\PPoi$, its spatial part is given by
$A_S=\Gamma_\infty d_S\Gamma_\infty^\inv$ and condition
\eqref{boundary} is invariant under \bea \label{rinvariance}
\Gamma_\infty \mapsto\Gamma_\infty \cdot w^\inv, \eea where  $w$
is now  an arbitrary $\PPoi$-valued function of $x^0$. We
therefore propose to  promote $w$ to a dynamical variable and to
couple  it $w$ to $\mu$ and $s$ via the kinetic term \bea
 -\langle \mu J_0+sP_0, w^\inv \partial_0 w\rangle.
\eea To ensure gauge invariance, we combine this term
 with the standard
kinetic term $\langle \mu J_0+sP_0, h^\inv \partial_0 h\rangle$,
so that the total kinetic term for the variables $h,w,\mu,s$
associated to the distinguished puncture is given by \bea
\label{kinpun} - \langle \mu J_0+sP_0, w^\inv \partial_0 w\rangle
+\langle \mu J_0+sP_0, h^\inv
\partial_0 h\rangle. \eea Defining a Poincar\'e element
$T\in\PPoi$ as in \eqref{boundel} and setting $g=hw^\inv$, we can
rewrite  \eqref{kinpun}  as \bea \label{cotakin} \langle
T,g\partial_0g^\inv \rangle, \eea which shows in particular that
the kinetic term for the variables associated to the puncture does
not depend on the way the coefficient of the curvature singularity
\eqref{singcoeff} is parametrised. Moreover, the expression
\eqref{cotakin} has a simple geometric interpretation. As
explained in \cite{we4}, this kinetic term is derived from the
symplectic potential of the canonical symplectic structure on the
cotangent bundle of $\PPoi$.


\subsection{Action and equations of motion}

We now couple the kinetic term \eqref{kinpun} for the
distinguished puncture  to the Chern-Simons action
\eqref{ordaction} and show that solutions of the resulting
equations of motion have the required behaviour near the
distinguished puncture representing spatial infinity. We define
our action for spacetimes $M\approx\RR\times\surfb$  as
\begin{align}
 &S[A_S,A_0, h_i, \mu,s, h,w]=\label{actiongrav}
\tfrac{1}{2}\int_\RR dx^0\int_{\surfb}\langle\partial_0
A_S\wedge A_S\rangle\\
 &-\int_\RR dx^0 \sum_{i=1}^n \langle \mu_i J_0+s_i P_0\,,
\, h_i^\inv\partial_0 h_i\rangle +\int_\RR dx^0
\langle \mu J_0+s P_0, h^\inv\partial_0 h\rangle\nonumber\\
 &+\int_\RR dx^0\!\!\int_{\surfb}\!\!\!\langle A_0, F_S +
 h(\mu J_0+s P_0)h^\inv\delta^{(2)}(x-\xb)dx^1\wedge dx^2\!-\!\sum_{i=1}^n
 T_i\delta^{(2)}(x-x_{(i)})dx^1\wedge dx^2 \rangle\nonumber\\
&-\int_\RR dx^0 \langle \mu J_0+sP_0, w^\inv \partial_0 w\rangle.
\nonumber
\end{align}
Varying the action \eqref{actiongrav} with respect to $A_0$
 we obtain the constraint
\bea \label{a0vary} F_S(x)= -h(\mu J_0+s
P_0)h^\inv\delta^{(2)}(x-x_\infty)+\sum_{i=1}^n
T_i\delta^{(2)}(x-x_{(i)}). \eea Variation with respect to $A_S$
and $h_i$ yields the equations \eqref{asvary} and \eqref{hivary}.
It remains to derive the equations of motion from the variation of
the new variables:
\begin{align}
&\delta\mu \quad\Rightarrow\quad\langle h^\inv A_0(x_0,\xb)h+h^\inv\partial_0 h-w^\inv\partial_0 w, J_0\rangle=0,\label{muvary}\\
&\delta s \quad\Rightarrow\quad\langle h^\inv A_0(x_0,\xb)h+h^\inv\partial_0 h-w^\inv\partial_0 w, P_0\rangle=0,\label{svary}\\
&\delta h\quad\Rightarrow\quad [\mu J_0+s P_0, h^\inv A_0(x_0,\xb) h+ h^\inv\partial_0 h]=\partial_0\mu J_0+\partial_0 s P_0,\label{hvary}\\
&\delta w\quad\Rightarrow\quad [\mu J_0+s P_0, w^\inv\partial_0
w]=\partial_0 \mu J_0+\partial_0 s P_0\label{wvary}.
\end{align}
The last equation implies
\begin{align}
\label{musconst} &\partial_0 \mu=\partial_0
s=0\\
\label{wtimeconst}&[w^\inv\partial_0 w, J_0]=[w^\inv\partial_0 w
,P_0]=0.
\end{align}
Mass $\mu$ and spin $s$ of the asymptotic cone are therefore
constants of motion. Furthermore, combining equations
\eqref{muvary}, \eqref{svary}, \eqref{hvary}, we find that the
function $A_0$ at the puncture is given by
\begin{align}
\label{a0equ}
A_0(x_0,\xb)=h w^\inv\partial_0 w h^\inv+h\partial_0 h^\inv=
hw^\inv\partial_0(wh^\inv).
\end{align}


To see that these equations capture the geometry of a spinning cone
 in the region $\RR \times P_\infty$ we first derive a general
 formula for  gauge fields that solve the equations of
 motion in $\RR\times P_\infty$.

\begin{theorem}
For any gauge field solving the equations of motion \eqref{a0vary}
to \eqref{wvary} there exists a region $P_\infty$ containing the
distinguished puncture and with the topology of a punctured disc
in which the gauge field can be written as
\bea \label{solution}
 A=\Gamma d \Gamma^\inv,
\eea with a multi-valued trivialising function $\Gamma: \RR\times
P_\infty\rightarrow\PPoi$ of the form
\begin{align}
\label{gravttrivi} \Gamma= \gamma_\infty
(e^{\frac{\mu\varphi}{2\pi} J_0},\tfrac{s\varphi}{2\pi} P_0)
w^\inv,
\end{align}
where $\gamma_\infty: \RR\times P_\infty\rightarrow\PPoi$,
$\gamma_\infty(x_0,\xb)=h$, $w:\RR\rightarrow\PPoi$ satisfies
\eqref{wtimeconst} and $\mu,s$ are real constants.
\end{theorem}

{\bf Proof:}

Choosing a neighbourhood $P_\infty$ of the distinguished puncture
with the topology of a punctured disc, we can trivialise any gauge
field as in \eqref{solution} with a multi-valued function
$\Gamma:\RR\times P_\infty\rightarrow \PPoi$, which is unique up
to right-multiplication with a constant Poincar\'e element. It is
therefore sufficient to show that the function $\Gamma$ defined in
\eqref{gravttrivi} yields a solution of the equations of motion
near the distinguished puncture. For this, we note that the
spatial component of the gauge field $A_S=\Gamma d_S \Gamma^\inv$
for the trivialising function \eqref{gravttrivi} is given by
 \bea
\label{compgfas} A_S= -\gamma_\infty (\tfrac{\mu}{2\pi}
J_0+\tfrac{s}{2\pi} P_0)\gamma_\infty^\inv d\varphi +\gamma_\infty
d\gamma_\infty^\inv, \eea which, as discussed above, is the most
general form of a  gauge field satisfying  the curvature
constraint \bea \label{again} F_S(x_0,\xb)=-h( \mu J_0+s P_0
)h^\inv \delta^{(2)}(x-x_\infty)dx^1\wedge dx^2. \eea The
component $A_0$ near the puncture is given by
 \bea
 \label{compgfa0}
A_0=\Gamma\partial_0\Gamma^\inv=\gamma_\infty
\partial_0\gamma_\infty^\inv+\gamma_\infty (e^{\frac{\mu\varphi}{2\pi}
J_0}, \tfrac{s\varphi}{2\pi} P_0) w^\inv\partial_0
w(e^{-\frac{\mu\varphi}{2\pi} J_0}, \tfrac{-s\varphi}{2\pi}
P_0)\gamma_\infty^\inv, \eea and taking into account
\eqref{wtimeconst} we find \bea \label{compgfa02}
 A_0=\gamma_\infty \partial_0\gamma_\infty^\inv+
\gamma_\infty  w^\inv\partial_0 w\gamma_\infty^\inv. \eea It
follows that equation \eqref{asvary} is satisfied and at the
distinguished puncture we have \eqref{a0equ}\bea \label{a0w}
 A_0(x_0,x_\infty)=
h\partial_0 h^\inv+ h w^\inv \partial_0 w h^\inv=  hw^\inv
\partial_0( w h^\inv),\eea
which proves the claim.\hfill $\Box$

The equations of motion derived from the action \eqref{actiongrav}
are therefore equivalent to the requirement that the gauge field
takes the form \eqref{solution}, \eqref{gravttrivi} in the region
$\RR\times P_\infty$. To establish a link with the gauge field
\eqref{dreib2} for the spinning cone, we now consider the
$x^0$-evolution of $w$.  The general solution of
\eqref{wtimeconst} is \bea \label{wsol} w(x^0)=\hat w c(x^0) \eea
where $\hat w $ is a constant element in $\PPoi$ and $c$ is an
$x^0$-dependent element of the abelian subgroup $T_\gothc$.
 Parametrising $c$ as
\bea \label{cpara} c= (e^{\frac{\mu\varphi_0}{2\pi}J_0},
(t_0+\tfrac{s\varphi_0}{2\pi})P_0), \eea with $\varphi_0$ and $t_0$ functions
of $x^0$,
the gauge field \eqref{solution} on $\RR \times
P_\infty$ takes the form
\begin{align}
\label{3dgfcm} A=\gamma_\infty( \tfrac{\mu}{2\pi}d
(\bphi+\varphi_0) J_0
+ (dt_0+\tfrac{s}{2\pi} d(\bphi+\varphi_0)P_0 )\gamma_\infty^\inv+\gamma_\infty
d\gamma_\infty^\inv,
\end{align}
where we have again used $\bphi =2\pi-\varphi$. Up to a trivial
shift of the angle, \eqref{3dgfcm} agrees with \eqref{boundgf3},
and we recover the gauge field \eqref{dreib2} for a spinning cone
by setting
\bea \label{gaugeee}
\gamma_\infty=\left(1,-\frac{r}{1-\tfrac{\mu}{2\pi}}\cos(\bphi +\varphi_0)
P_1-\frac{r}{1-\tfrac{\mu}{2\pi}}\sin(\bphi+\varphi_0) P_2\right),
\eea where
$r$ is a function of $\rho$ only. The function $r$ can be chosen
to increase to an arbitrary large  value as $\rho$ decreases, but
we must have $r=0$  at the distinguished puncture in order for
\eqref{gaugeee} to be well-defined ($\bphi$-independent) there.

We have thus shown that the equations of motion derived from the
action \eqref{actiongrav} have solutions which, in a suitably
gauge, have the  geometrical interpretation  of a spinning cone
near the distinguished puncture representing spatial infinity. As
explained in Sect.~\ref{physmot}, the Poincar\'e element
$w\in\PPoi$ specifies the embedding of the cone into Minkowski
space.  Equations \eqref{wtimeconst}, \eqref{wsol} which restrict
the $x^0$-evolution of $w$ to right-multiplication with elements
of the abelian  subgroup $T_\gothc$ therefore imply that the
embedding of the cone's symmetry  axis does not change. The
right-multiplication with $x^0$-dependent elements of $T_\gothc$
corresponds to rotations around  and translation in the direction
of the cone's axis. We will see in Sect.~\ref{gaugetrafo} that
they are related to gauge transformations. In accordance with the
remarks made at the end of  Sect.~\ref{physmot} they only acquire
a physical significance after gauge fixing.

The fact that the $x^0$-evolution of $w$ is restricted to
right-multiplication with elements of $T_\gothc$ allows
one in particular to restrict
 $w$ in the action \eqref{actiongrav} to the abelian  subgroup
 $T_\gothc$,
 i.~e.~to set $\hat w =0$ in \eqref{wsol}.
Geometrically, this  amounts to restricting ourselves to
embeddings of the cone into  Minkowski space such that the axis of
the cone coincides with the $T$-axis, i.e. to   centre of
mass frames of  the universe. With the parametrisation
\eqref{wsol} and \eqref{cpara}
 we find
\begin{align}
\label{cmterm} -\langle\mu J_0+s P_0,w^\inv\partial_0
w\rangle=-\mu
\partial_0 t_0-\tfrac{\mu s}{2\pi}\partial_0
\varphi_0-\partial_0\left(\tfrac{\mu s \varphi_0}{2\pi}\right).
\end{align}
Inserting this expression into the action \eqref{actiongrav} and
dropping
 the total time derivative , we then obtain
the centre of mass action
\begin{align}
&S[A_S,A_0, h_i, \mu,s, h,w]=\label{cmaction}\tfrac{1}{2}\int_\RR dx^0\int_{\surfb}\langle\partial_0 A_S\wedge A_S\rangle\\
 &
 -\int_\RR dx^0 \sum_{i=1}^n \langle \mu_i J_0+s_i P_0\,,\, h_i^\inv\partial_0 h_i\rangle +\int_\RR dx^0 \langle \mu J_0+s P_0, h^\inv\partial_0 h\rangle\nonumber\\
 &+\int_\RR dx^0\!\!\int_{\surfb}\!\!\!\langle A_0, F_S +
 h(\mu J_0+s P_0)h^\inv\delta^{(2)}(x-\xb)dx^1\wedge dx^2\!-\!\sum_{i=1}^n  T_i\delta^{(2)}(x-x_{(i)})dx^1\wedge dx^2 \rangle\nonumber\\
&-\int_\RR dx^0 (\mu\partial_0 t_0 +\tfrac{\mu s}{2\pi}
\partial_0 \varphi_0 ).\nonumber
\end{align}
We should stress that $\varphi_0$ and $t_0$ are merely coordinates
used for parametrising elements of the abelian group $T_\gothc$.
They have a geometrical interpretation because the gauge field
takes the form \eqref{3dgfcm} in terms of them. They will be
useful when establishing a link with the metric formulation of
gravity in Sect.~\ref{metric}, but are not particularly convenient
for discussing gauge invariance and symmetry of our model. For
that purpose it is much better to work with the general Poincar\'e
element $w$,  with  the identification \eqref{phirange}.


\section{Symplectic structure}

\label{symp}

\subsection{Symplectic structure on the extended phase space}

After defining an action functional which implements spatial
infinity as a distinguished puncture, we will now investigate  a
description of the associated symplectic structure on the phase
space. The canonical symplectic form corresponding to the action
\eqref{actiongrav} is given by
\begin{align}
\label{hwdsympformgrav} \Omega=-&\delta\langle \mu J_0+s P_0, h^\inv\delta h\rangle+\delta\langle \mu J_0+s P_0,
w^\inv\delta w\rangle + \sum_{i=1}^n \delta\langle \mu_i J_0+s_i P_0,
h_i^\inv\delta h_i\rangle\nonumber\\
+&\tfrac{1}{2}\int_{\surfb} \langle \delta A_S\wedge\delta A_S\rangle.
\end{align}
It defines a symplectic structure on an auxiliary space,
parametrised by the variables $w,h,h_i\in\PPoi$, the particle's
masses $\mu_i$ and spins $s_i$, total mass $\mu$ and spin $s$ of
the universe and the spatial gauge field $A_S$. The physical phase
space $\mathcal P$ is obtained from this auxiliary space by
imposing the constraint \eqref{a0vary} on the curvature of the
spatial gauge field $A_S$, fixing the values of the particle's
masses $\mu_i$ and spins $s_i$ and dividing by the associated
gauge transformations. However, in practice it proves difficult to
parametrise the physical phase space explicitly and to express its
symplectic structure in terms of a set of independent parameters.
We therefore describe its symplectic structure in terms of a
two-form $\Omega$ on an extended phase space $\mathcal{P}_{ext}$,
on which most, but not all gauge freedom is eliminated.

\begin{figure}
\protect\input epsf \protect\epsfxsize=12truecm \protect
\centerline{\epsfbox{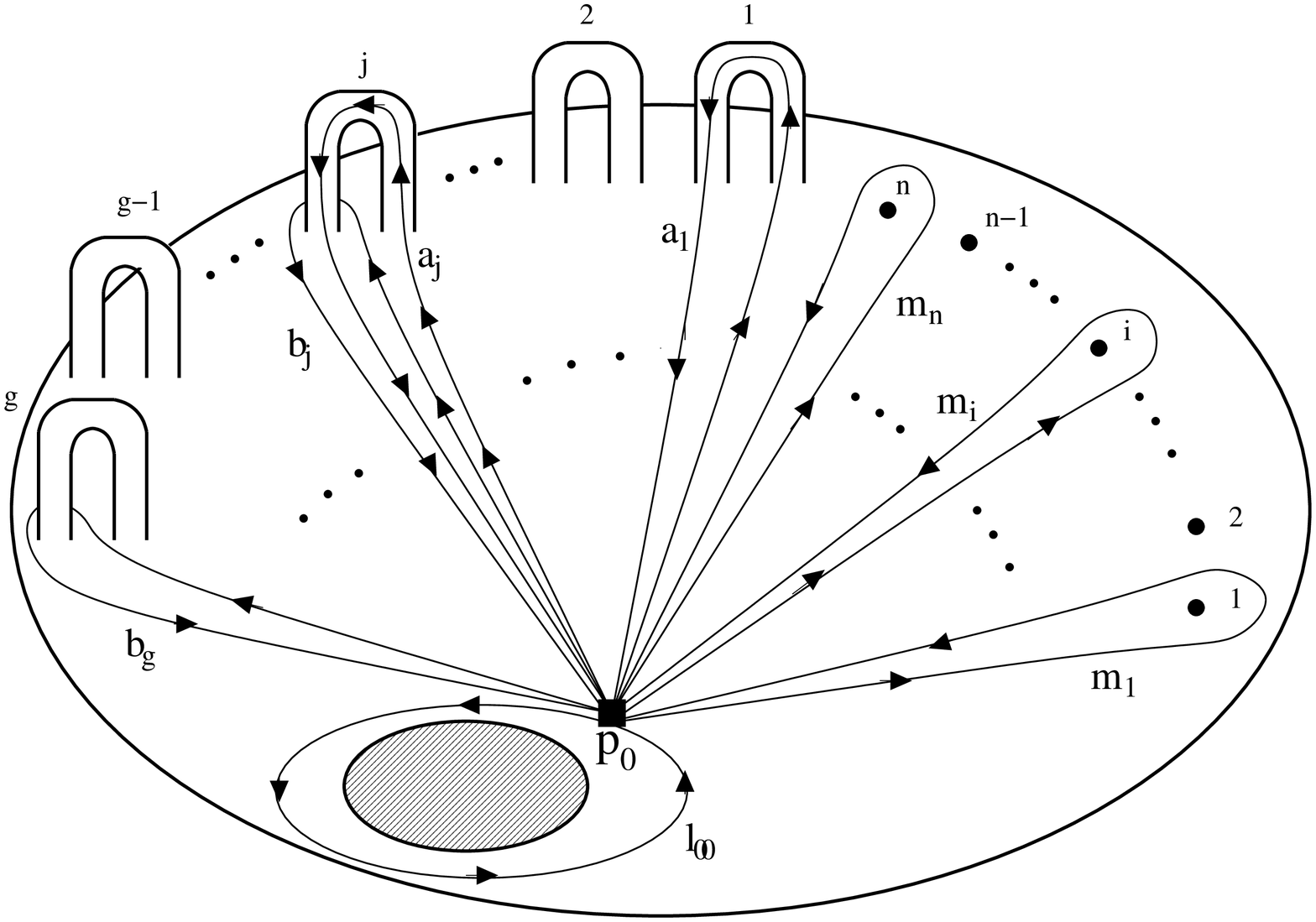}} \caption{ The generators of the
  fundamental group  $\pi_1(\surfb, \pp_0)$}
\label{pi1gen}
\end{figure}

Our description is based on the results in \cite{we4}, where we
applied the method introduced by Alekseev and Malkin \cite{AMII}
to derive such a description of the phase space for the general
case of Chern-Simons theory with gauge groups $\prgr$ on manifolds
$M\approx\RR\times\surfb$.  The action functional considered in
\cite{we4} with the choice $\prgr=\PPoi$ gives \eqref{actiongrav},
and we can therefore directly apply the results from \cite{we4}.
We then obtain a parametrisation of the phase space in terms of
the total mass $\mu$ and spin $s$ of the universe, the particle's
masses and spins $\mu_i,s_i$ and the Poincar\'e element
$w\in\PPoi$ together with $n+2g$ $\PPoi$-valued variables closely
related to the holonomies of a set of generators of the
fundamental group $\pi_1(\surfb)$. These variables, in the
following referred to as holonomy variables, are obtained as
follows. One chooses an arbitrary base point $p_0\in\surfb$ on the
spatial surface and considers a set of curves $\mi,\aj,\bjj$,
$i=1,\ldots,n$, $j=1,\ldots, g$ based at  $p_0$ whose homotopy
equivalence classes generate the fundamental group $\pi_1(\surfb,
\pp_0)$, see Fig.~\ref{pi1gen}.
As their homotopy equivalence classes generate the fundamental
group freely, the holonomies  $\aj,\bjj$ associated to the curves
$a_j,b_j$ around the handles are general elements of the group
$\PPoi$, while the holonomies $M_i$ lie in fixed conjugacy classes
\bea
\label{cclasses}
\mathcal{C}_{\mu_i s_i}=\{g_i(e^{-\mu_i J_0}, -s_i P_0)g_i^\inv |
g_i \in\PPoi\}
\eea
 determined by the masses  and spins of the
particles. The holonomy $L_\infty$ of the curve $l_\infty$ around
the distinguished puncture with respect to the base point $p_0$ is
given as a function of the group elements $\mi,\aj,\bjj$ by \bea
\label{boundaryh} L_\infty^\inv=[B_g,A_g^\inv]\cdots[B_1,
A_1^\inv]\cdot\mf\cdots\me, \eea and can be parametrised in terms
of the mass $\mu$ and spin $s$ of the universe and a general
element $g_\infty\in\PPoi$ as
\begin{align}
\label{inftyhol}
L_\infty^\inv=g_\infty (e^{-\mu J_0}, -s P_0)g_\infty^\inv.
\end{align}
Note that the element $g_\infty\in\PPoi$ in the parametrisation
\eqref{inftyhol} is not unique and determined only up to
right-multiplication with elements of the abelian  subgroup $T_\gothc$
\begin{align}
\label{gredund}
g_\infty\mapsto g_\infty (e^{-\psi J_0}, \tau P_0)\quad\Rightarrow\quad L_\infty^\inv\mapsto L_\infty^\inv.
\end{align}
It will become apparent in Sect.~\ref{gaugetrafo} that such
transformations are related to gauge transformations acting on the
extended phase space. We now introduce another set of
$\PPoi$-valued variables $\tilde\mi$,$\tilde\aj$, $\tilde\bjj$
obtained by conjugating the holonomies $\mi,\aj,\bjj$ with
$g_\infty^\inv$
\begin{align}
\label{tildeholdef}
&\tilde \mi=g_\infty^\inv \mi g_\infty & &\tilde \aj=g_\infty^\inv \aj g_\infty & &\tilde \bjj=g_\infty^\inv \bjj g_\infty.
\end{align}
From this definition and  \eqref{inftyhol} it follows that these variables are subject to the relation
\begin{align}
\label{tilderel}
[\tilde B_g,\tilde A_g^\inv]\cdots[\tilde B_1,\tilde A_1^\inv]\cdot\tilde\mf\cdots\tilde\me=(e^{-\mu J_0}, -s P_0).
\end{align}
 We parametrise these holonomy variables as in \eqref{holpara} in
 terms of Lorentz transformations $\tilde v_\mi,\tilde u_\aj,\tilde
 u_\bjj\in\LLor$ and angular momenta
 $\tilde\pbj_\mi,\tilde\pbj_\aj,\tilde\pbj_\bjj\in\RR^3$ according to
\begin{align}
\label{holparam0}
&\tilde \mi=(\tilde u_\mi,-\Ad(\tilde u_\mi)\tilde \pbj_\mi)=(\tilde v_\mi,\tilde \bx_\mi)(e^{-\mu_i J_0},-s_i P_0)(\tilde v_\mi,\tilde \bx_\mi)^\inv\\
&\tilde \ai=(\tilde u_\ai,-\Ad(\tilde u_\ai)\tilde \pbj_\ai)\nonumber\\
&\tilde \bi=(\tilde u_\bi,-\Ad(\tilde u_\bi)\tilde \pbj_\bi)\nonumber,
\end{align}
so that for the particles
 \bea \label{holparm0} \tilde
u_\mi=\tilde v_\mi e^{-\mu_i J_0}\tilde
v_\mi^\inv\qquad\tilde\pbj_\mi=(1-\Ad(\tilde u_\mi^\inv))\tilde
\bx_\mi+\Ad(\tilde v_\mi)s_iP_0. \eea The constraint
\eqref{partconstraints} then takes the form \bea
\label{spinconstr} \langle\Ad(\tilde v_\mi)
J_0,\tilde\pbj_\mi\rangle-s_i=0. \eea In terms of these variables,
our extended phase space $\mathcal{P}_{ext}$ can be characterised
as follows
\begin{theorem} (Theorem 6.3 in \cite{we4})
Consider the extended phase space $\mathcal{P}_{ext}$ parametrised
by the holonomy variables $\tilde v_\mi,\tilde u_\aj,\tilde
u_\bjj\in\LLor$,
$\tilde\pbj_\mi,\tilde\pbj_\aj,\tilde\pbj_\bjj\in\RR^3$,
$w\in\PPoi$, $\mu,s\in\RR$ and the two-form $\Omega$
\begin{align}
\label{finresultgrav} \Omega=\delta\left(  \langle \mu J_0 +s P_0,
w^\inv\delta w\rangle-\tfrac{1}{2}\mu \delta s +\Theta(\tilde
v_\mi,\tilde\pbj_\mi, \tilde u_\aj,\tilde \pbj_\aj, \tilde
u_\bjj,\tilde\pbj_\bjj)\right),
\end{align}
where $\Theta$ is the  one-form
\begin{align}
\label{gravtheta}
&\Theta(\tilde v_\mi, \tilde u_\aj, \tilde u_\bjj, \tilde \pbj_\mi, \tilde \pbj_\aj, \tilde \pbj_\bjj)=\\
&\;\;\;\sum_{i=1}^n \langle\; \delta(\tilde u_{M_{i-1}}\cdots
\tilde u_{M_1})(\tilde u_{M_{i-1}}\cdots \tilde u_{M_1})^\inv-
\delta \tilde v_\mi \tilde v_\mi^{-1}\,,\,
\tilde \pbj_\mi\rangle\nonumber\\
&+\sum_{i=1}^g\langle\; \delta(\tilde u_{H_{i-1}}\cdots \tilde
u_{H_1}\tilde u_\mf\cdots\tilde u_{M_1})(\tilde u_{H_{i-1}}\cdots
\tilde
u_{H_1}\tilde u_\mf\cdots\tilde u_{M_1})^\inv\,,\, \tilde \pbj_\ai\rangle\nonumber\\
&\qquad-\langle\; \delta(\tilde u_\ai^\inv \tilde u_\bi^\inv
\tilde u_\ai \tilde u_{H_{i-1}}\cdots \tilde u_{H_1}\tilde
u_\mf\cdots\tilde u_{M_1})(\tilde u_\ai^\inv \tilde u_\bi^\inv
\tilde u_\ai \tilde u_{H_{i-1}}\cdots \tilde u_{H_1}\tilde
u_\mf\ldots\tilde u_{M_1})^\inv\,,\,
\tilde \pbj_\ai \rangle\nonumber\\
&+\sum_{i=1}^g\langle\; \delta(\tilde u_\ai^\inv \tilde u_\bi^\inv
\tilde u_\ai \tilde u_{H_{i-1}}\cdots \tilde u_{H_1}\tilde
u_\mf\cdots\tilde u_{M_1})(\tilde u_\ai^\inv \tilde u_\bi^\inv
\tilde u_\ai \tilde u_{H_{i-1}}\cdots \tilde u_{H_1}\tilde
u_\mf\cdots\tilde u_{M_1})^\inv\,,\,
\tilde \pbj_\bi \rangle\nonumber\\
&\qquad-\langle\; \delta( \tilde u_\bi^\inv \tilde u_\ai \tilde
u_{H_{i-1}}\cdots \tilde u_{H_1}\tilde u_\mf\cdots\tilde u_{M_1})(
\tilde u_\bi^\inv \tilde u_\ai \tilde u_{H_{i-1}}\cdots \tilde
u_{H_1}\tilde u_\mf\cdots\tilde u_{M_1})^\inv\,,\, \tilde \pbj_\bi
\rangle\nonumber
\end{align}
with  $\tilde u_{H_i}=[\tilde u_{\bi},\tilde u_\ai^\inv]=\tilde
u_\bi \tilde u_\ai^\inv \tilde u_\bi^\inv \tilde u_\ai$.
The
physical phase space is obtained from $\mathcal{P}_{ext}$ by
imposing the constraint \eqref{tilderel} and the $n$ spin
constraints \eqref{spinconstr} and by dividing by the associated
gauge transformations. The two-form $\Omega$ restricted to the
constraint surface  is the pull-back of
the symplectic structure on the physical phase space to the constraint
surface.
\end{theorem}

We will now discuss the physical interpretation of this result in the context of (2+1)-dimensional gravity. For this, we
 parametrise the element $w\in\PPoi$ as
\begin{align}
\label{embedpar} w=(v, \bx)(1, \tfrac{1}{2} s P_0)=(v,
\bx+\tfrac{1}{2}s\Ad(v)P_0)\qquad\text{with}\;v\in\LLor,\bx\in\RR^3,
\end{align}
so that
\begin{align}
\label{wderivhelpid}
w^\inv\delta w=(v^\inv\delta v,\Ad(v^\inv)\delta\bx)+(0, \tfrac{1}{2}\delta s P_0)+(0,[v^\inv\delta v,\tfrac{1}{2}sP_0 ]).
\end{align}
The last term gives no contribution to the pairing of $w^\inv\delta w$ with $\mu J_0+s P_0$, and the contribution of the second term cancels with the term $-\frac{1}{2}\mu\delta s$ in the two-form \eqref{finresultgrav}. We can therefore rewrite the two-form \eqref{finresultgrav} as the derivative of a one-form
\begin{align}
\label{physgrav} \Omega=\delta\left(  \langle \bp\,,\, \delta\bx\rangle +
\langle s P_0, v^\inv\delta v\rangle +\Theta(\tilde
v_\mi,\tilde\pbj_\mi, \tilde u_\aj,\tilde \pbj_\aj, \tilde u_\bjj,\tilde\pbj_\bjj)\right).
\end{align}
with the momentum three-vector
\begin{align}
\label{pinftydef}
\bp=\mu\Ad(v)J_0.
\end{align}

We see that the one-form in \eqref{finresultgrav} and
\eqref{physgrav} is  the sum of the one-form  $\Theta$
depending  only on the holonomy
 variables $\tilde v_\mi,\tilde u_\aj,\tilde u_\bjj\in\LLor$, $\tilde\pbj_\mi,\tilde\pbj_\aj,\tilde\pbj_\bjj\in\RR^3$ and of terms involving only the
 variables $\mu,s$ and $(v,\bx)\in\PPoi$ (equivalently $w\in\PPoi$)
 associated to  spatial infinity.
 The former describes  the
relative motion of the different particles and handles in a
centre of mass frame of the universe.
 The  terms which depend only on the variables associated to spatial
 infinity describe the motion of a distinguished
centre of mass frame of the universe
with respect to an observer (or, equivalently, the motion of an
observer relative to a distinguished centre of mass frame of the universe).
 The geometrical interpretation  of
 the Poincar\'e transformation $w=(v,\bx+\tfrac{1}{2} s\Ad(v) P_0)$,
explained in Sects.~\ref{particle} and \ref{physmot}, is that it
specifies the embedding of the spinning cone into
Minkowski space.
In particular it maps  the $T$-axis
 to the axis of the embedded cone. Thus the
two Poincar\'e elements $w$ and  $(v,\bx)$ are
equivalent in the sense that the associated embeddings differ only
by a  shift $T\mapsto T+\tfrac{s}{2}$ along the $T$-axis.

Our description in terms of the extended phase space with the
two-form $\Omega$ therefore decouples the relative motion of
different particles and handles and  the
centre of mass motion relative to an observer.
The holonomy variables  $\tilde v_\mi,\tilde
u_\aj,\tilde u_\bjj\in\LLor$,
$\tilde\pbj_\mi,\tilde\pbj_\aj,\tilde\pbj_\bjj\in\RR^3$ describing
the relative motion are linked to the mass $\mu$ and spin $s$ of
the asymptotic cone by the constraint \eqref{tilderel}. We can
view this constraint as the gravitational equivalent of
the usual condition specifying the total  energy of a system of
free particles as a function of the parameters of their relative
motion.


\subsection{Gauge transformations and symmetries}
\label{gaugetrafo}

The physical phase space is obtained from the extended phase space
$\mathcal{P}_{ext}$ with variables $\tilde v_\mi,\tilde
u_\aj,\tilde u_\bjj\in\LLor$,
$\tilde\pbj_\mi,\tilde\pbj_\aj,\tilde\pbj_\bjj\in\RR^3$,
$\mu,s\in\RR$ and  $(v,\bx)\in\PPoi$ and two-form \eqref{physgrav}
by imposing the constraints \eqref{tilderel} and
\eqref{spinconstr} and dividing by the associated gauge
transformations. We now give explicit expressions for the action
of these gauge transformations on the extended phase space.

In a framework where the symplectic structure of the physical phase space is described in terms of a two-form $\Omega$ on an extended phase space $\mathcal{P}_{ext}$,
gauge invariance manifests itself in degeneracy of $\Omega$ on the constraint surface $\mathcal{C}\subset \mathcal{P}_{ext}$.
There exist vector fields $Z$ on the constraint surface, the infinitesimal generators of the gauge transformations,
such that the contraction of the two-form $\Omega$ with these vector fields is zero on the constraint surface $\iota_Z\Omega=0$.
 In the situation at hand, where  $\Omega=\delta\chi$ is given as the
exterior derivative of a one-form $\chi$ as in \eqref{finresultgrav},
\eqref{physgrav} and where
the gauge transformations are group actions on the extended phase space, there is a rather straightforward way of determining these gauge
transformations
 and their generators directly from the one-form. The method is discussed in detail in \cite{we4}, Sect.~6, and can be summarised as follows.

Suppose we have the action $\rho$
of a one-parameter group $h(\epsilon)$, $h(0)=1$,
 on $\cal C$, and $Z$ is the vector field generated by
$\rho (h(\epsilon))$. Then $Z$ is the infinitesimal generator of a
gauge transformation if
 \bea \label{gaugecond}
\iota_Z\delta\chi =0 \Leftrightarrow {\cal L}_Z\chi
-\delta\iota_Z\chi =0. \eea
To check if this condition holds we consider the change of $\chi$
under pull-back with $\rho( h(\epsilon))$, where $\epsilon$ is allowed
to  be a  function on $\cal C$.
If  the
one-form $\chi$ changes according to
\bea
\label{gaugetest} \chi \mapsto \chi +\delta(\epsilon H) +{\cal
O}(\epsilon^2) \eea with a function $H$ on the constraint surface,
it follows that $\iota_{Z}\chi=\chi (Z)=H$ for the associated
vector field $Z$  and ${\cal L}_{Z}\chi =\delta H$. Hence, the
group action $\rho(h(\epsilon))$ is a gauge transformation if and
only if the one-form $\chi$  transforms according to
\eqref{gaugetest}.

We now apply this method to our extended phase space with
the two-form  \eqref{physgrav}. The first gauge symmetry we
encounter is related to the $\PPoi$-valued  constraint
\eqref{tilderel} and  arises from the redundancy in parametrising
the action \eqref{actiongrav} in terms of the Poincar\'e elements
$h,w\in\PPoi$. It is easy to see that the \eqref{actiongrav} is
invariant under simultaneous right-multiplication of $w$ and $h$
by an element of the abelian subgroup $T_\mathfrak{c}$, while all
other variables stay fixed. On the extended phase space, such a
transformation manifests itself in simultaneous
right-multiplication of $w$ or, equivalently, $(v,\bx)$ and
conjugation of all holonomy variables
\begin{align}
\label{boundgt}
&(v,\bx)\mapsto (v,\bx) (e^{-\Phi J_0}, T P_0)\\
&\tilde X\mapsto (e^{-\Phi J_0}, T P_0)\tilde X (e^{\Phi J_0}, - T
P_0)\qquad\forall\tilde X\in\{\tilde M_1,\ldots,\tilde B_g\} \nonumber\\
&\tilde v_\mi\mapsto e^{-\Phi J_0}\tilde v_\mi.\nonumber
\end{align}
This transformation preserves the constraint \eqref{tilderel} and
the spin constraints \eqref{spinconstr} for the particles.
As explained in Sect.~\ref{physmot},  the right multiplication of $(v,\bx)$
with $(e^{-\Phi J_0}, T P_0)$ corresponds to the  rotation
around  and time
translations in the direction of the asymptotic cone's symmetry axis.
As anticipated  also in       Sect.~\ref{physmot} such transformations
have no physical meaning if we simultaneously rotate and
time-translate  all other  coordinates used in the description of
the universe. This is expressed mathematically  in the gauge invariance of our
symplectic structure   under \eqref{boundgt}.

In order to prove that \eqref{boundgt} is a gauge transformation
we first
determine the transformation behaviour of the two-form
\eqref{physgrav}. Noting  that
$(v,\bx)\mapsto (v,\bx) (e^{-\Phi J_0}, T P_0)$ is equivalent to
$w\mapsto w(e^{-\Phi J_0}, T
P_0)$ we find that under the transformation \eqref{boundgt}
\begin{align}
\langle (\mu J_0,s P_0), w^\inv\delta w\rangle\mapsto &\langle
 (\mu J_0,s P_0), w^\inv\delta w\rangle+\mu\delta T-s\delta\Phi.
\end{align}
Furthermore, it follows from Lemma 6.2 in \cite{we4}, but can also
verified by direct calculation from definitions \eqref{holparm0}
and \eqref{gravtheta}, that under \eqref{boundgt} the one-form
\eqref{gravtheta} transforms according to
\begin{align}
\Theta(\tilde v_\mi,\tilde\pbj_\mi, \tilde u_\aj,\tilde
\pbj_\aj, \tilde u_\bjj,\tilde\pbj_\bjj)&\mapsto
 \Theta(\tilde v_\mi,\tilde\pbj_\mi, \tilde u_\aj,\tilde
\pbj_\aj, \tilde u_\bjj,\tilde\pbj_\bjj)+s\delta\Phi+ T\delta\mu.
\end{align}
The sum
$\Theta +\langle (\bmu,\bs), w^\inv\delta w\rangle$ therefore changes by the
exact one-form $\delta(\mu T)$ and the transformation \eqref{boundgt} is a gauge transformation.

Another set of gauge transformations is generated by the spin
constraints \eqref{spinconstr} for  the particles which reflect
the redundancy in the parametrisation \eqref{holparm0} of $\tilde
u_\mi$ in terms of $\tilde v_\mi$. The elements $\tilde u_\mi$ are
invariant under right-multiplication of the corresponding elements
$\tilde v_\mi$ by a rotation of angle $\Phi_i\in\RR$ \bea
\label{spingt} \tilde v_\mi\mapsto \tilde v_\mi e^{-\Phi_i
J_0}\qquad\Phi_i\in\RR. \eea From formula \eqref{gravtheta}, we
find that this transformation changes  the one-form
$\Theta$ according to \bea \label{partrotgt} \Theta(\tilde
v_\mi,\tilde\pbj_\mi, \tilde u_\aj,\tilde \pbj_\aj, \tilde
u_\bjj,\tilde\pbj_\bjj)\mapsto
 \Theta(\tilde v_\mi,\tilde\pbj_\mi, \tilde u_\aj,\tilde
\pbj_\aj, \tilde u_\bjj,\tilde\pbj_\bjj)+\sum_{i=1}^n s_i\delta \Phi_i.
\eea
On the constraint surface, where  the spins $s_i$ of the particles are all fixed, the last term in \eqref{partrotgt} is a total derivative, and we see that the transformations \eqref{spingt} are indeed gauge transformations.

The extended phase space is therefore given by $6(n+2g)$ parameters
associated to the holonomy variables in the one-form $\Theta$, $6n$ parameters in the Poincar\'e element $w\in\PPoi$ together with the mass $\mu$ and spin $s$ of the universe. These parameters are subject to six constraints from the condition \eqref{tilderel} and $n$ spin constraints \eqref{spinconstr} for the particles. Furthermore, there are the two gauge transformations \eqref{boundgt} as well as $n$ gauge transformations of the form \eqref{partrotgt}. The physical phase space is obtained by imposing the constraints and dividing by  the gauge transformations. It therefore has dimension
\bea
\label{dofcount}
6(n+2g)+6+2-(6+n)-(2+n)=4n+12g.
\eea
In particular, for the case $g=0$, i.~e.~a system of $n$ particles on a disc, the dimension of the phase space is $4n$, the same as for a system of $n$ free particles without gravitational interaction.

In contrast to the gauge symmetries, physical symmetries are given by group actions $\rho(h(\epsilon))$ on the constraint surface $\mathcal C$, which depend only
 on a {\em parameter},  not a function $\epsilon$, which leave the two-form $\Omega$ invariant  and  which commute with the gauge symmetries.
 The last condition ensures that they give rise to group actions on the physical phase space obtained by taking the quotient with the gauge transformations.
 (For a detailed discussion of the general relation between symmetries on an extended phase space, gauge symmetries and physical symmetries see \cite{giu2}.)
  In the case at hand, the physical symmetries are  a global action on the (2+1)-dimensional Poincar\'e group associated to spatial infinity. On the extended phase space, this Poincar\'e group action is given by
\begin{align}
\label{wsymm}
(v,\bx)\mapsto g (v,\bx)\qquad g\in\PPoi
\end{align}
or, equivalently, $w\mapsto gw$. As the gauge symmetry
\eqref{boundgt} acts on $(v,\bx)$ and $w$ by right-multiplication
and the gauge symmetry \eqref{spingt} leaves these variables
invariant, we see that the transformation \eqref{wsymm}  indeed
commutes with all gauge symmetries. Furthermore, as the parameters
in the group action \eqref{wsymm} are not dependent on the phase
space variables $\delta g=0$, it preserves the term $w^\inv\delta
w$. Using expression \eqref{finresultgrav} for the two-form
$\Omega$ , we find that  it leaves $\Omega$  invariant and
therefore gives rise to a symmetry acting on the physical phase
space. This symmetry action reflects the asymptotic Poincar\'e
symmetry of our spacetime. It describes a Poincar\'e
transformation \eqref{rinvariance} acting on the embedding
Minkowski space, i.~e.~a Poincar\'e transformation with respect to
the distinguished reference frame associated to spatial infinity.


\subsection{Centre of mass frames}

As discussed in Sect.~\ref{physmot}, the centre of mass frames of
the universe, where the axis of the cone coincides with the $T$-axis
in the embedding Minkowski space, are  defined by the
restriction of the element $w\in\PPoi$ to the abelian  subgroup $T_\gothc$,
i.~e.~a combination of a rotation and a time translation.
Using again the parametrisation \eqref{cpara} for such elements,
but incorporating  a trivial shift in
the time coordinate $t_0\mapsto t_0-\frac{1}{2}s$, we set
\begin{align}
\label{cmwcond}
(v, \bx)=(e^{\frac{\mu\varphi_0}{2\pi}J_0}, (t_0+\tfrac{s}{2\pi}\varphi_0)).
\end{align}
In the centre of mass frames, the momentum $\bp$ in
\eqref{physgrav} is directed along the $T$-axis and, up to a total
derivative  $\delta(\frac{\mu s }{2\pi}\varphi_0)$,  the two-form
\eqref{physgrav} reduces to
\begin{align}
\label{cmsymp}
\Omega=\delta\chi(\tilde v_\mi,\tilde u_\aj,\tilde u_\bjj,\tilde\pbj_\mi,\tilde\pbj_\aj,\tilde\pbj_\bjj, \mu,s,t_0,\varphi_0)
\end{align}
with
\begin{align}
\label{cmpot} \chi(\tilde v_\mi,\tilde u_\aj,\tilde
u_\bjj,\tilde\pbj_\mi,\tilde\pbj_\aj,\tilde\pbj_\bjj,
\mu,s,t_0,\varphi_0)=\Theta(\tilde v_\mi,\tilde u_\aj,\tilde
u_\bjj,\tilde\pbj_\mi,\tilde\pbj_\aj,\tilde\pbj_\bjj)+\mu\delta
t_0+\tfrac{\mu s}{2\pi}\delta\varphi_0,
\end{align}
which is the one-form  we would have obtained by
 working directly with action \ref{cmaction}.

While the gauge transformations  \eqref{spingt} for each particle
take the same form in a centre of mass frame, the gauge
transformations \eqref{boundgt} now  correspond to a shift in the
time parameter $t_0$ and the angle $\varphi_0$
\begin{align}
\label{cmgt}
&t_0\mapsto t_0+\tau\qquad\varphi_0\mapsto\varphi_0+\psi \\
&\tilde X\mapsto (e^{\frac{\mu\psi}{2\pi} J_0}, (\tau+\tfrac{s}{2\pi}\psi) P_0)\tilde X (e^{-\frac{\mu\psi}{2\pi} J_0}, -(\tau+\tfrac{s}{2\pi}\psi)\qquad\forall\tilde X\in\{\tilde M_1,\ldots,\tilde B_g\}\nonumber\\
&\tilde v_\mi\mapsto e^{\frac{\mu\psi}{2\pi} J_0}\tilde v_\mi.\nonumber
\end{align}
Written in terms of $\varphi_0$ and $t_0$ these
gauge transformations look more complicated than \eqref{boundgt},  but
their  interpretation is precisely  as explained  after
\eqref{boundgt},
see also the discussion in the next subsection.


\subsection{Gauge fixing and physical degrees of freedom}

\label{gaugefix}
After deriving an explicit description of the phase space structure
in terms of  a   two-form
 on the extended phase space, we will now discuss how to eliminate the gauge freedom in our description by imposing appropriate gauge fixing conditions.

We start by considering the gauge transformations \eqref{spingt}
associated to the spin constraints \eqref{spinconstr} for the
particles. These gauge symmetries reflect the redundancy in the
parametrisation \eqref{holparm0}, which can
in principle
be eliminated by requiring that the elements $\tilde v_\mi$ are pure boosts, parametrised by an angle $\theta_\mi\in[0,2\pi)$ specifying the direction and a boost parameter $\beta_\mi\in\RR^+_0$
\bea
\label{spinfix}
\tilde v_\mi=e^{-\theta_\mi J_0} e^{-\beta_\mi J_2} e^{\theta_\mi J_0}.
\eea
In practice, it is more convenient to work with the redundant
parametrisation \eqref{holparm0} and to keep track of resulting
gauge transformations.

We now turn to the gauge transformations \eqref{boundgt}
associated to the constraint \eqref{tilderel} which relates the
holonomy variables to mass $\mu$ and spin $s$ of the asymptotic
cone. As discussed in Sects.~\ref{physmot}
and \ref{gaugetrafo}, these gauge transformations
 reflect the fact that the centre of mass condition requiring the axis of
the asymptotic cone to coincide with the $T$-axis only
determines a coordinate frame up to rotations and time
translations. Rotating and time translating both, the holonomy
variables describing the relative motion of different particles
and handles with respect to an observer and the Poincar\'e element
relating the coordinate frame of the observer to a fixed frame
where the axis of the cone coincides with the $T$-axis as in
\eqref{boundgt} yields an equivalent description of the same
physical state. In physical terms, the situation can be described
as follows. The observables associated to spatial infinity allow
an observer to specify a distinguished  reference frame. Once such a
frame is specified,  motions {\em with respect to this frame} are
physical. However, applying  the same transformation to {\em both}
the distinguished frame {\em and} all other observable objects in the
universe does not yield a different state of the universe but a
description of the same state in terms of a different coordinate
system.

It is useful to again consider the corresponding situation in
(3+1)-dimensions. There, one would impose  as a boundary condition
that the metric is asymptotically the Minkowski metric,
and a distinguished
frame associated to spatial infinity would usually be defined by
referring to a set of very distant fixed stars. The physical
information about a system of particles in such a
(3+1)-dimensional universe would then be given by its motion
relative to these fixed stars. However, applying the same
Poincar\'e transformation to the distinguished  frame {\em and} the
motion of the particles would not yield a different state  but
only a gauge equivalent description of the same state in terms of
a different coordinate system. The situation in (2+1)-dimensions
is analogous, only that instead of the Poincar\'e symmetries
associated to Minkowski space, such gauge transformations take
values in the symmetry group of the asymptotic cone, the abelian
subgroup $T_\mathfrak{c}$.

From the discussion above, it is clear that there are two possible
ways of gauge fixing. The first is to fix a distinguished
reference frame defined by the boundary, i.~e. to impose a
restriction on the Poincar\'e element $(v,\bx)\in\PPoi$ \bea
\label{timerotfix} (v,\bx)=(e^{-\theta_\infty J_0}
e^{-\beta_\infty J_2} e^{\theta_\infty J_0}\!\!,\; x^1
P_1+x^2P_2)\,,\,\quad
\theta_\infty\in[0,2\pi),\beta_\infty\in\RR^+_0. \eea A
corresponding condition for the centre of mass case would be to
set the variables $\varphi_0$ and $t_0$ in \eqref{cmpot} to zero.
However, the gauge fixing condition \eqref{timerotfix} has the
disadvantage that is not compatible with the asymptotic Poincar\'e
symmetry \eqref{wsymm} and yields to complicated expressions for
the centre of mass motion.

An alternative gauge fixing condition without these problems is
obtained by defining a reference frame in terms of
 the motion of one of the particles.
For instance, assuming that not all particles are at rest
relative to the centre of mass of the universe,
we can impose without loss of generality
 that the first particle moves along
the $1$-axis,
which amounts to setting $\theta_\me=0$ in
\eqref{spinfix}
\bea \label{lortrfix} \tilde
v_\me=e^{-\beta_\me
J_2}\qquad \beta_\me\in\RR^+_0,
\eea
and demand that its angular
momentum three-vector is given by
\bea \label{jtrfix}
\tilde \pbj_\me=\Ad(e^{-\beta_\me J_2})s_1 P_0+ j_\me^2 P_2,
\eea which
amounts to a restriction on the particle's position vector
$\tilde \bx_\me$.
Although in principle it is possible to use the gauge
fixing \eqref{timerotfix}, \eqref{lortrfix} and \eqref{jtrfix} to
express the two-form $\Omega$  in terms of independent
parameters, there  is no obvious way of doing this in practice
 for a general genus $g$ surface with arbitrarily many
massive, spinning particles.
 For the case of a  genus $g=0$
surface with $n>1$ massive, but spinless particles, we will see in
Sect.~\ref{metric} that
 there
is a way of solving the constraints which leads to a fairly simple
expression for the one-form $\chi$. Spatial surfaces
$S_{0,2}^\infty$ with two massive, spinning particles are
discussed further in Sect.~\ref{quantcond}. In this case, the
motion of the two particles is determined completely by the mass
$\mu$ and spin $s$ of the universe and the Poincar\'e element $w$.

The centre of mass motion is governed by the symplectic potential
\bea
 \label{cmterm2} \langle(\mu J_0+s P_0)\,,\,w^\inv\delta w
\rangle.
\eea
After the imposition of the gauge fixing conditions
\eqref{lortrfix}, \eqref{jtrfix}, the Poincar\'e element
$w\in\PPoi$ becomes a physically meaningful observable, namely a
Poincar\'e transformation with respect to a fixed frame in which
the axis of the cone coincides with the $T$-axis. As the
ambiguity in the definition of $w$ is removed by the gauge fixing
conditions \eqref{lortrfix}, \eqref{jtrfix}, right-multiplication
of this element by rotations and time shifts
\bea
\label{hphystr}
w\mapsto w\cdot(e^{-\Phi J_0}, T P_0)
\eea
now represents
rotations and time translations with respect to this reference
frame. As discussed in Sect.~\ref{particle}, the angle coordinate $\Phi$ and
time coordinate $T$ obey  the identification
\bea
\label{identify}
(T,\Phi)\sim(T+s, \Phi+2\pi-\mu)
\eea
leading to the identification  \eqref{phirange} for the
$\PPoi$-element
 $w$. We will
see in Sect.~\ref{quantcond} that this condition gives rise to a
quantisation condition on mass $\mu$ and spin $s$ of the universe.


\subsection{Relationship to the flower algebra}
\label{flower}

To end this section,  we briefly explain the relationship between
the symplectic structure defined by the two-form
\eqref{finresultgrav} or, equivalently, \eqref{physgrav} and the
Poisson structure we gave in \cite{we1}.
 As explained in the introduction, that Poisson structure was 
not derived from a field-theoretical treatment of  the  boundary.
Instead, we argued on symmetry grounds that a Poisson structure 
defined by Fock and Rosly in \cite{FR} as an auxiliary Poisson
structure in the study of  moduli spaces of flat connections
on a surface without boundary  could be used to describe the 
Poisson structure of the 
phase space of an open universe in the Chern-Simons formulation of 
(2+1)-dimensional gravity.
 In \cite{we2} we called
the Poisson algebra corresponding to that Poisson structure
the   flower algebra, and we adopt
this terminology here.
The Poisson structure given in \cite{we1}
is defined on the space $\left(\PPoi\right)^{2g+n}$, and it is
symplectic when restricted to the space \bea \label{leaf}
\mathcal{C}_{\mu_1 s_1}\times\ldots\times \mathcal{C}_{\mu_n s_n}
\times \left(\PPoi\right)^{2g} \eea spanned by the holonomy
variables $M_1,\ldots, M_n,A_1,B_1,\ldots,A_g,B_g$. Parametrising
them via
\begin{align}
\label{holparrr}
& \mi=( u_\mi,-\Ad( u_\mi)\bj_\mi)=( v_\mi,  \bx_\mi)(e^{-\mu_i
  J_0},-s_i P_0)
( v_\mi,\bx_\mi)^\inv\\
&\ai=(u_\ai,-\Ad( u_\ai)\bj_\ai)\nonumber\\
&\bi=( u_\bi,-\Ad(u_\bi) \bj_\bi)\nonumber,
\end{align}
so that
 \bea
\label{holpa} u_\mi=v_\mi e^{-\mu_i J_0} v_\mi^\inv  \qquad
\bj_\mi=(1-\Ad( u_\mi^\inv)) \bx_\mi+\Ad(v_\mi)s_iP_0, \eea the
symplectic structure on the symplectic leaf \eqref{leaf} is most
easily described  by pulling it back from the $\LLor$-holonomies
$u_{M_1},\ldots,u_{M_n}$ to the $\LLor$-elements
$v_{M_1},\ldots,v_{M_n}$. As shown in \cite{we3} that pull back is
the two-form \bea \Omega_{\text{\tiny flower}}=\delta \Theta(
v_\mi,\bj_\mi,  u_\aj, \bj_\aj,  u_\bjj, \bj_\bjj), \eea with
$\Theta$ defined as in \eqref{gravtheta}. To compare
$\Omega_{\text{\tiny flower}}$
  with $\Omega$ in \eqref{physgrav} we express $\Omega_{\text{\tiny
    flower}}$ in terms of the holonomies \eqref{tildeholdef}
and the $\PPoi$-element $g_\infty$. Parametrising
$g_\infty=(v_\infty,\bx_\infty)$ and
\bea
g_\infty(e^{-\mu J_0}, -s
  P_0)g_\infty^\inv=(u_\infty,-\Ad(u_\infty)\bj_\infty)
\eea
we have the equivalent formula
\bea
\label{flowersymp}
\Omega_{\text{\tiny flower}}= \delta\left(  -\langle
\delta u_\infty  u_\infty^\inv\,,\,\bx_\infty \rangle -
\langle s P_0, v_\infty ^\inv\delta v_\infty \rangle +\Theta(\tilde
v_\mi,\tilde\pbj_\mi, \tilde u_\aj,\tilde \pbj_\aj, \tilde u_\bjj,\tilde\pbj_\bjj)\right).
\eea
As stated in the introduction,  the last term is the same as the
term describing the relative
 motion of handles and particles in the universe
in \eqref{physgrav}. The  first
two terms are analogous to the first two terms in \eqref{physgrav}
describing the centre of mass motion of the universe. However,
while our discussion in Sect.~4 has provided a clear interpretation
of the Poincar\'e element  $w$,  
we have no such interpretation for
the Poincar\'e element  $g_\infty$ because the symplectic structure
\eqref{flowersymp}  was  not derived from a field-theoretical treatment
of the boundary. We note, however,
that  we can  impose the analogue of the 
centre of mass  condition  by requiring 
$g_\infty$  to take values in the abelian subgroup
$T_\gothc$. Parametrising
such as elements as
\bea
\label{flowercm}
g_\infty=
(e^{-\frac{\mu\varphi_0}{2\pi}J_0},
(-t_0-\tfrac{s\varphi_0}{2\pi})P_0),
\eea
we obtain agreement with our symplectic structure \eqref{cmsymp}
for centre of mass frames.


\section{The link with the metric formulation}
\label{metric}

Due to the different role of degenerate dreibeins in the two
theories, the relation between the phase space of
(2+1)-dimensional gravity in its Chern-Simons and its Einstein
formulation is rather subtle and  not yet fully clarified
\cite{Matschull2}. In particular, it is not obvious how our
description of the phase space for open universes based on the
Chern-Simons formalism is related to other approaches which take
the metric viewpoint. In this section, we show that our
description of the phase space agrees with the results obtained by
Matschull \cite{Matschull1} who derives a description of the phase
space for open universes of genus $g=0$ with $n$ massive, but
spinless particles in a centre of mass frame of the universe.

The derivation in \cite{Matschull1} is based entirely on the
metric formulation of (2+1)-dimensional gravity and relies on an
ADM-decomposition of the spacetime. Spatial infinity is
incorporated as a boundary with the boundary condition that
dreibein and spin connection asymptotically take the form
\eqref{dreib2} associated to a spinning cone. The formalism makes
use of  graph embedded into a spatial slice of the spacetime and
describes the phase space by means of a symplectic potential on an
extended phase space of link variables. For the convenience of the
reader, we summarise the results of \cite{Matschull1} in the
Appendix and show how they can be specialised to a minimal graph.

To establish   a link between our description of the phase space
and  the work \cite{Matschull1}, we consider manifolds
$M\approx\RR\times S_{0,n}^\infty $, where the spatial surface  is
a disc with $n$ massive, but spinless particles and restrict
attention to centre of mass frames.  In that case, the variables
parametrising the extended phase space $\mathcal{P}_{ext}$ are the
total mass $\mu$ and spin $s$ of the universe, together with the
time and angle coordinates $t_0$, $\varphi_0$ and the holonomy
variables $\tilde v_\me,\ldots,\tilde v_\mf$,
$\tilde\pbj_\me,\ldots,\tilde\pbj_\mf$. In terms of these,
 the symplectic
potential on $\mathcal{P}_{ext}$ in the centre-of-mass frame
is given by \eqref{cmpot},
\eqref{gravtheta}
\begin{align}
\label{origpot} \chi(\tilde v_\mi, \tilde \pbj_\mi, \mu,s,
t_0,\varphi_0) =&\Theta(\tilde v_\mi,\tilde \pbj_\mi)+\mu\delta
t_0+\tfrac{\mu s}{2\pi}\delta\varphi_0.
\end{align}
As discussed in Sect.~\ref{symp}, the variables are subject to a
spin constraint \eqref{spinconstr}  for each of the particles,
where now $s_i=0$, and the constraint \eqref{tilderel} with
associated gauge transformations \eqref{spingt} and
\eqref{boundgt}.

It turns out that for the case of $n$ spinless particles on a
disc, some of these constraints can be solved explicitly by
parametrising the extended phase space in terms of a new set of
variables. We have the following lemma, which can be proved by
straightforward calculation using the identity $\delta \tilde
u_\mi \tilde u_\mi^\inv=\delta \tilde v_\mi \tilde v_\mi^\inv-
\tilde u_\mi \delta \tilde v_\mi \tilde v_\mi^\inv \tilde
u_\mi^\inv$.

\begin{lemma} $\quad$
\label{matlemma}

1. For genus $g=0$ and $n$ spinless particles, the one-form
 \eqref{gravtheta}
 takes the form
\begin{align}
\label{pot1} \Theta(\tilde u_\mi,\tilde\bx_\mi)=&\sum_{i=1}^n
\langle \delta(\tilde u_{M_{i-1}}\cdots \tilde u_\me)(\tilde
u_{M_{i-1}}\cdots \tilde u_\me)^\inv-\delta (\tilde
u_{M_{i}}\cdots \tilde u_\me)(\tilde u_{M_{i}}\cdots \tilde
u_\me)^\inv\,,\,\tilde \bx_\mi\rangle
\end{align}
in terms of the  variables $\tilde u_\mi,\tilde\bx_\mi$ related to
$\tilde v_\mi$, $\tilde\pbj_\mi$ by
\begin{align}\label{holparm}  \tilde u_\mi=\tilde v_\mi e^{-\mu_i J_0}\tilde
v_\mi^\inv\qquad \tilde\pbj_\mi=(1-\Ad(\tilde
u_\mi^\inv))\tilde\bx_\mi.
\end{align}

2. Introducing a  new set of variables
via
\begin{align}
\label{gdef2} &g_1=\tilde u_\me^\inv,\;\; g_i=\tilde
u_\me^\inv\cdots \tilde u_\mi^\inv &  &i=2,\ldots,n-1\\
\label{zdef2} &\bz_i=\tilde \bx_{M_{i+1}}-\tilde \bx_\mi &
&i=1,\ldots, n-1,
\end{align}
we have the alternative expression
\begin{align}
\label{endthetam20} \Theta(g_i,\bz_i, \tilde\bx_\mf)
=&-\sum_{i=1}^{n-1}\langle g_i^\inv \delta
g_i,\bz_i \rangle-\langle \delta(\tilde u_\mf
g_{n-1}^\inv)g_{n-1}\tilde u_\mf^\inv\,,\,\bx_\mf\rangle.
\end{align}
\end{lemma}

In terms of the variables $g_i$, $\bz_i$, $\tilde u_\mf$ and
$\tilde\bx_\mf$,  the  constraint \eqref{tilderel} takes the form
\begin{align}
\tilde M_n\cdots\tilde M_1=(\tilde u_\mf g_{n-1}^\inv, -\Ad(\tilde
u_\mf g_{n-1}^\inv)\tilde\pbj_{K_n})\approx(e^{-\mu J_0}, -s P_0)
\end{align}
with \bea \label{mangmom} \tilde \pbj_{K_n}=(1-\Ad(\tilde g_{n-1}
u_\mf^\inv))\,\tilde \bx_\mf-\sum_{i=1}^{n-1} (1-\Ad(g_i))\,\bz_i,
\eea or, equivalently,
\begin{align}
&\tilde u_\mf g_{n-1}^\inv\approx e^{-\mu J_0}\label{finhols}\\
&\label{finspin} \sum_{i=1}^{n-1}\langle (1-\Ad(g_i))\bz_i\,,\,J_0\rangle
\approx -s\\
&(1-\Ad(g_{n-1} \tilde u_\mf^\inv))\tilde \bx_\mf-\sum_{i=1}^{n-1}
(1-\Ad(g_i))\bz_i + \sum_{i=1}^{n-1}\langle(1-\Ad(g_i))\bz_i,
J_0\rangle P_0\approx 0.\label{finspinortho}
\end{align}
By adding a function of the constraint \eqref{finhols}, we can
rewrite the symplectic potential \eqref{origpot} as
\begin{align}
\label{endthetam2} \chi(g_i,\bz_i, \tilde\bx_\mf, \mu,s,
t_0,\varphi_0 )=&-\sum_{i=1}^{n-1}\langle g_i^\inv \delta
g_i,\bz_i \rangle+\mu\delta t_0+\tfrac{\mu
s}{2\pi}\delta\varphi_0+\langle \tilde \bx_\mf, J_0\rangle
\delta\mu.
\end{align}

As this expression does not depend on the variable $\tilde u_\mf$
and only depends on the component of $\tilde\bx_\mf$ parallel to
$P_0$, we can now solve the constraints \eqref{finhols} and
\eqref{finspinortho} by setting
 \begin{align}
&\label{fixlor}u_\mf=g_{n-1}e^{-\mu J_0}\\
&\label{xnfixs} \tilde \bx_\mf-\langle \tilde \bx_\mf,
J_0\rangle P_0=\frac{1}{1-\Ad(e^{\mu
J_0})}\sum_{i=1}^{n-1}(1-\Ad(g_i))\bz_i -\langle
(1-\Ad(g_i))\bz_i\,,\,J_0\rangle P_0,
\end{align}
where $\frac{1}{1-\Ad(e^{\mu J_0})}$ denotes the inverse of the
bijective map $(1-\Ad(e^{\mu J_0}))|_{\text{Span}(P_1,
P_2)}:$\linebreak $\text{Span}(P_1, P_2)\rightarrow
\text{Span}(P_1, P_2)$. Furthermore, we can remove the component
$\tilde\bx_\mf$ parallel to $P_0$ by shifting our time parameter
\bea \label{timedef} T_0=t_0-\langle \tilde
\bx_\mf\,,\,J_0\rangle. \eea This allows one to eliminate the
variables $\tilde u_\mf$, $\tilde \bx_\mf$ from the extended phase
space and, after adding a total derivative, to rewrite the
symplectic potential \eqref{endthetam2}  as
\begin{align}
\label{endthetam3} \chi(g_i,\bz_i, \mu,s,T_0,\varphi_0)=&
-T_0\delta \mu +\tfrac{\mu
s}{2\pi}\delta\varphi_0-\sum_{i=1}^{n-1}\langle g_i^\inv \delta
g_i\,,\,\bz_i \rangle.
\end{align}

Our extended phase space is then parametrised by $n-1$ Lorentz
transformations $g_i$, $n-1$ vectors $\bz_i\in\RR^3$ and the four
real parameters $\mu,s,T_0,\varphi_0$. While the constraints
\eqref{finhols} and \eqref{finspinortho} are no longer present,
these variables are still subject to the constraint
\eqref{finspin}, and we have traded the spin constraints
\eqref{spinconstr}  for a  mass constraint for each particle. With
the definitions $g_n=e^{\mu J_0}$, $g_i^\inv g_{i-1}=e^{-\tilde p_\mi^a
J_a}$ these mass constraints read
\begin{align}
\label{finmassconst} &\sqrt{\tilde \bp_\mi^2}-\mu_i\approx 0\qquad
 i=1,\ldots,n.
\end{align}
Using the methods from Sect.~\ref{gaugetrafo}, we find that the
gauge transformations associated to the constraint \eqref{finspin}
are given by
\begin{align}
\label{spinunivgt}
&\varphi_0\mapsto \varphi_0-\tfrac{2\pi}{\mu}\psi\;,\quad
T_0\mapsto T_0+\tfrac{s}{\mu}\psi
\\
&g_i\mapsto e^{-\psi J_0}g_i e^{\psi J_0}\;,\quad
\bz_i\mapsto\Ad(e^{-\psi J_0}))\bz_i,
\quad
i=1,\ldots,n-1,\nonumber
\end{align}
and the gauge transformations corresponding to the mass
constraints \eqref{finmassconst} are
\begin{align}
\label{minmassgt2}
C_1:\qquad &\bz_1\mapsto\bz_1-\tfrac{\tau_1}{\mu_1} {\tilde\bp}_\me \\
C_i:\qquad &\bz_i\mapsto \bz_i-\tfrac{\tau_i}{\mu_i} {\tilde\bp}_\mi,\quad\bz_{i-1}\mapsto \bz_{i-1}+\tfrac{\tau_i}{\mu_i}{\tilde\bp}_\mi\qquad i=2,\ldots,n-1\nonumber\\
C_n:\qquad&\bz_{n-1}\mapsto
\bz_{n-1}+\tfrac{\tau_n}{\mu_n}\tilde\bp_\mf,\quad T_0\mapsto
T_0-\tfrac{\tau_n}{\mu_n}\,\langle {\tilde p}_\mf^aJ_a,
P^0\rangle\nonumber,
\end{align}
with parameters $\psi,\tau_i\in\RR$. The physical phase space is
obtained from the $6(n-1)+4$-dimensional extended phase space with
symplectic potential \eqref{endthetam3} by imposing the $n+1$
first-class constraints \eqref{finspin}, \eqref{finmassconst} and
dividing by the associated gauge transformations
\eqref{spinunivgt}, \eqref{minmassgt2} and therefore has the
dimension
\begin{align}
6(n-1)+4-2(n+1)=4n-4.
\end{align}

This description of the physical phase space in terms of the
extended phase space with variables $g_i,\bz_i$, $i=1,\ldots,n-1$
and parameters $\mu,s,T_0,\varphi_0$ and symplectic potential
\eqref{endthetam3} is exactly the one obtained by specialising the
formalism in \cite{Matschull1} to a minimal graph, for details see
Sect.~\ref{minsect} in the appendix. Our description of open
universes from the Chern-Simons viewpoint, where spatial infinity
is treated as a distinguished puncture therefore agrees with the
results in \cite{Matschull1} for a system of $n$ spinless
particles on a disc in a centre of mass frame of the universe.
While the physical interpretation of the results is more readily
apparent from the metric viewpoint, our formalism is more general.
It gives gives an efficient description of spacetimes of general
genus $g$ with an arbitrary number of particles with non-vanishing
spin and without restriction to centre of mass frames.  In
particular, it is not clear how the formalism in \cite{Matschull1}
can be extended to include spinning particles, while the inclusion
of spin poses no problems in the Chern-Simons framework.

\section{The Quantisation condition on mass and spin}

\label{quantcond}

After giving a description of the phase space of (2+1)-dimensional gravity for open universes in terms of the two-form $\Omega$ on an extended phase space, we will now demonstrate how this description and the results in \cite{we2} give rise to a quantisation condition on the total mass $\mu$ and spin $s$ of the universe.

It is shown in Sect.~\ref{symp} that the eight degrees of freedom
associated to the centre of mass are the mass $\mu$ and spin $s$
of the universe and the Poincar\'e element $w=(v, \bx)(1,
\tfrac{1}{2} s P_0)\in\PPoi$. Defining $\bp$ as in
\eqref{pinftydef}, setting $\hat\bp=\tfrac{1}{\mu}\bp$
 and adding a total derivative, we can write the symplectic potential \eqref{cmterm2} associated to the centre of mass motion as
\begin{align}
\label{finpot}
\langle (\mu J_0+s P_0)\,,\, w^\inv\delta w\rangle-\tfrac{1}{2}\mu\delta s&=\langle\bp,\delta\bx\rangle+\langle s P_0, v^\inv\delta v\rangle=
\langle \bk\,,\, \delta v v^\inv\rangle-\delta\mu\langle \hat{\bp}\,,\,\bx\rangle\\
&=\langle \bk\,,\, \delta v v^\inv\rangle- T\delta\mu,\nonumber
\end{align}
where the vector $\bk$ is defined in terms of $\bx$, $\bp$ and $s$
as in  \eqref{porbit}
 and
\bea
T=\langle\hat\bp\,,\,\bx\rangle.
\eea
From the identification \eqref{identify} it follows that
these variables are subject to the identification \bea
\label{lastident} (T,\mu, \bk,v)\sim(T+s, \mu,\bk, v
e^{-(2\pi-\mu)J_0}). \eea

Expression \eqref{finpot} for the symplectic potential in terms of
the variables $T,\mu,v,\bk$ implies that the only non-vanishing
Poisson brackets of these variables are given by
\begin{align}
\label{kpbrac2}
&\{k_a, f(v)\}=\frac{d}{d\ee}f(e^{-\ee J_a}v)\;\,\forall f\in\cif(\LLor)\\
\label{kkbrac2} &\{ k_a,k_b\}=\ee_{ab}^{\;\;\;\;c} k_c\\
&\label{mutbrac2}\{\mu, T\}=-1.
\end{align}
The spin of the universe can be expressed in terms of the angular
momentum three-vector $\bk\in\RR^3$ and the Lorentz transformation
$v\in\LLor$ as \bea \label{lastspindef} s=\langle \bk\,,\,\Ad(v)
J_0\rangle, \eea and it follows from the brackets \eqref{kpbrac2},
\eqref{mutbrac2} that it generates a right-multiplication of the
Lorentz element  $v$ by a rotation
\bea
\label{rotation3}
s:\quad v\mapsto v e^{-\ee J_0}.
\eea
However, the
identification \eqref{lastident} implies that there is no value of
the flow parameter $\ee$ for which the flow in phase space
generated by the spin is equal to the identity. Instead, we find
that the generator of a rotation whose flow by $2 \pi$  is equal
to the identity is the alternative angular momentum
\begin{align}
\label{truerotgen2} J=(1-\frac{\mu}{2\pi})s,
\end{align}
which generates the transformation \bea \label{jflow2} J:\quad
(T,\mu,\bk,v)\mapsto(\,T+\ee\tfrac{s}{2\pi},\mu,\bk, ve^{-\ee
(1-\frac{\mu}{2\pi}) J_0}). \eea The phase space function $s$ and
$J$  and their relation was first discussed in the metric
formalism in a centre of mass frame in \cite{Matschull1} and
\cite{LM1}.

From expressions \eqref{kpbrac2}, \eqref{kkbrac2} and \eqref{mutbrac2}
 for the Poisson bracket, it is apparent that the Poisson algebra describing
the centre of mass degrees of freedom is of a special type, namely
the  semidirect product of the universal enveloping algebra of the   Lie
algebra $so(2,1)\oplus\RR$
with  the space of $\cif$-functions on $\LLor\times\RR$.
We  considered the quantisation of Poisson algebras of that type
 in our paper \cite{we2}. We can therefore apply our quantisation procedure
developed there to construct the quantum algebra and investigate its
irreducible Hilbert space representations. From Theorem 3.1 in \cite{we2}, it
follows that the quantum algebra for the Poisson algebra \eqref{kpbrac2}, \eqref{kkbrac2} and \eqref{mutbrac2} is the algebra
\begin{align}
\label{lastalg} \hat{\gothf}=U(so(2,1)\oplus\RR)\tenltimes \cif(\LLor\times\RR)
\end{align}
with multiplication
\begin{align}
\label{genmultlast}
&((\bk_1, T_1)\otimes F_1)\cdot((\bk_2,T_2)\otimes F_2)\\
&=((\bk_1,T_1)\!\cdot_U\!(\bk_2,T_2))\otimes F_1 F_2+i\hbar\, (\bk_2,T_2)
\otimes F_1 \{(\bk_1, T_1), F_2\}\nonumber\\
&\bk_1,\bk_2\in so(2,1),  F_1,F_2\in\cif( \LLor\times\RR,\mathbb{C}),\nonumber
\end{align}
where the components of the angular momentum three-vector $\bk$
and the time $T$ are identified with the generators of the Lie
algebra $ so(2,1)\oplus\RR$ so that $T_1=t_1 T$, $T_2=t_2 T$ with
$t_1,t_2\in\RR$. The Lorentz transformation $v$ and mass $\mu$
appear as the arguments of the functions
$F\in\cif(\LLor\times\RR)$:
\bea \{(\bk,T),
F\}(v,\mu)\!=\!\frac{d}{d\ee}|_{\ee=0}F(e^{-\ee k^a J_a}\cdot v ,
\mu\!+\!\ee)\; \qquad\forall \bk\in so(2,1),
F\in\cif(\LLor\times\RR).
\eea

As discussed in \cite{we2}, Sect.~3,
 the quantum algebra \eqref{lastalg} is represented on the space \bea \label{lastrepsp}
V=\cif_0(\LLor\times\RR), \eea
 of $\cif$-functions with compact support on the group $\LLor\times\RR$ according to
\begin{align}
\label{lastrep}
&\Pi(\bk)\psi(v,\mu)=i\hbar
\frac{d}{d\ee}|_{\ee=0}\psi(e^{-\ee k^a J_a }\cdot v,\mu)\\
&\Pi(T)\psi(v,\mu)=i\hbar  \frac{\partial}{\partial \mu }\psi (v,\mu )\nonumber\\
&\Pi(F)\psi(v,\mu)= F(v,\mu)\psi(v,\mu)\qquad\nonumber\forall \bk\in so(2,1),T\in\RR, F\in\cif(\LLor\times\RR).
\end{align}
In particular,  the mass operator $\mu$ of the universe  is identified
with the function
\bea
\mu\in\cif(\LLor\times\RR): \;(v,\tilde
\mu)\mapsto\tilde \mu
\eea
 and acts by multiplication \bea \label{massactlast}
 \Pi(\mu)\psi(v,\mu)=\mu \psi(v,\mu).
\eea
Furthermore, it follows from
\eqref{lastspindef} and \eqref{lastrep} that the spin operator $s$ acts by right-multiplica\-tion of the argument $v$ by an  infinitesimal
rotation
\begin{align}
\label{spinactlast} \qquad\Pi(s)\psi(v,\mu)= i\hbar \frac{d}{d\ee}|_{\ee=0}\psi(v e^{-\ee J_0},\mu),
\end{align}
and, consequently, the action of the angular momentum operator $J$ in
\eqref{truerotgen2} is given by
\bea
\label{jactlast2} \Pi(J)\psi(v,\mu)=i
\hbar \frac{d}{d\ee}|_{\ee=0}\psi(v \cdot e^{-\ee(1-\frac{\mu}{2\pi})
  J_0},\mu).
\eea

To derive a quantisation condition for mass and spin of the
universe, we decompose the Hilbert space \eqref{lastrepsp} into
eigenspaces $V_{\mu s}$ of the (commuting) mass and spin operators:
 \bea
\Pi(s)\psi=  s\;\psi\qquad\qquad
\Pi(\mu)\psi=\mu\psi\qquad\qquad\qquad\forall \psi\in V_{\mu s}.
\eea
From \eqref{jactlast2}, it then follows that these spaces are
also eigenspaces of the angular momentum operator $J$. The quantum
counterpart of the condition that the flow in phase space
generated by the angular momentum $J$ with flow parameter $2\pi$
is the requirement that the unitary operator $e^{2\pi
\frac{J}{i\hbar}}$ maps each state to itself. As this operator
acts on the eigenspaces $V_{\mu s}$ according to \bea \Pi(e^{2\pi
\frac{J}{i\hbar}})\psi=e^{i\frac{(2\pi-\mu)s}{\hbar}}\psi
\qquad\forall \psi\in V_{\mu s}, \eea this implies $(2\pi-\mu)s\in
2\pi \hbar\,\ZZ$ or, equivalently,  the quantisation
condition
\bea
\label{spinquant}
s=\frac{\hbar
n}{1-\tfrac{\mu}{2\pi}}\qquad\text{with}\qquad n\in \ZZ.
\eea

The condition \eqref{spinquant} has been found  by various authors
in more specialised settings. Extrapolating from the
study of  scattering of a test particle in the background of
fixed conical geometry,
 't~Hooft, Deser, Jackiw and de Sousa Gebert \cite{Hooft, DJ,
   desousaj1} conjectured the quantisation condition
\eqref{spinquant}
for the total mass
 and spin of a two-particle system.
 Carlip \cite{Carlipscat} showed that such a condition for the total
 mass and spin of two
spinless particles
can be obtained without relying on a semi-classical
approximation and is linked to the action of the braid group.
This is generalised to spinning particles
in \cite{bamus} where the condition \eqref{spinquant}
is obtained  from the condition that  the action
of the ribbon element of the quantum double $D(\LLor)$ is
identical to a rotation by $2\pi$.  The first derivation of
\eqref{spinquant}
from first principles is  given in
the work of Louko and Matschull \cite{LM2}
on the (2+1)-dimensional Kepler problem  and the paper
\cite{Matschull1}
on the phase space of
a system of $n$ massive, but spinless particles.
There it is shown that   \eqref{spinquant} holds
for the total mass and spin of, respectively, two and $n$ massive but
spinless particles in  the metric formulation of (2+1)-dimensional
gravity.  Our derivation shows that \eqref{spinquant} holds
quite generally for  arbitrary genus $g$  and  $n$ spinning particles.

Note  that in the case of an open universe of genus $g=0$ with two
massive,  spinning particles,
 the discussion in this section  amounts to a complete quantisation of the
 system.
According to equation \eqref{dofcount},
the phase space of such a two-particle system is eight-dimensional,
and the relative motion of the two particles is determined entirely
by the total mass and spin of the system, for details see the
discussion  in \cite{bamus}.
 After the application of the gauge fixing procedure
outlined in Sect.~\ref{gaugefix}, the phase space is
parametrised entirely by the variables $\mu, T,v,\bk$ with
the identification \eqref{lastident}, and its Poisson structure
is determined by the symplectic potential \eqref{finpot}.
The resulting quantum theory  is identical to the one
derived in \cite{bamus} by exploring the
analogy of particles in (2+1)-dimensional gravity with anyons.

\section{Outlook and conclusion}

Based on the treatment of spatial infinity as a distinguished
puncture, this paper contains an explicit description of the phase
space structure of the Chern-Simons formulation of
(2+1)-dimensional gravity in an open universe with vanishing
cosmological constant. We restrict attention to universes of
product topology $\RR \times S$, where $S$ is a two-dimensional
oriented manifold, but allow for $S$ to have arbitrary topology
and an arbitrary number of punctures.  We see generality and
explicitness as the main advantages of our approach and expect our
results  to be useful in the following ways.

The treatment of spatial infinity as a  distinguished puncture
provides the basis for modelling other physical situations. Here
we considered the boundary condition that the universe looks
asymptotically like a spinning cone.  In that case the holonomy
around the distinguished puncture is, in the terminology of
Sect.~2,  the exponential of an elliptic element of $iso(2,1)$
However, it is known that the total holonomy for two particles
which move very rapidly relative to each other is not of this type
\cite{Gott}. This can be accommodated in our formalism without
difficulty. In other contexts it may also be of interest to
consider several boundary components, which could be modelled in
our formalism by including several distinguished punctures.

Furthermore, we expect
 our results  to be useful in linking different approaches to
(2+1)-dimensional gravity, and to provide
 a unifying viewpoint for existing and
future investigations of the phase space structure. This includes
studies of closed universes, which can be thought as  the special
case $\mu=s=0$ of the open universes considered here (though the
discussion of gauge invariance and gauge fixing in Sect.~5 needs
to be adapted). In  this paper we  explained  the link between
Matschull's and our description of the phase space for $n$
spinless particles on a disc. It would also be interesting to
clarify the relation between
 't Hooft's polygon approach \cite{Hooft2}
and the Chern-Simons formulation of (2+1)-dimensional gravity.
This should be possible on the basis of our paper since  't
Hooft's approach  is closely related to Matschull's description of
the phase space in \cite{Matschull1}. In particular, one should be
able  to establish an explicit link between  our description of
the phase space and the coordinates and symplectic structure given
in the recent papers \cite{KL} and \cite{Kadar}, which treat the
case where $S$ is a closed surface of arbitrary genus without
punctures using the polygon approach.

Finally, we expect the current paper to be useful in investigating
 the classical dynamics
and the quantisation of (2+1)-dimensional
gravity.    The literature on these topics is vast and we shall make not
attempt to survey it here, referring the reader instead to
\cite{Carlipbook} for references.
 Rigorous and explicit
 results exist for special cases such
as $n$ particle dynamics on a genus $g=0$ surface or closed universes
of the form $\RR \times T^2$, where $T^2$ is the two-torus.
There are unifying themes, such as the appearance of the braid
group in the analysis of $n$ particle dynamics,
but a unified and systematic treatment is still lacking.
The current paper  provides the starting point for such a treatment
since our description of the phase space structure is general,
explicit and based on variables which are amenable to
a   physical interpretation.


\subsection*{Acknowledgements}   BJS acknowledges support through
an EPSRC advanced fellowship during the early stages of this project
and   thanks Thomas Strobl for
discussions about boundary conditions in field theory and gravity.

\appendix

\section{Matschull's description of the phase space in \cite{Matschull1}}

\subsection{General formalism}
In this appendix, we summarise the description of phase space and
symplectic structure given by Matschull \cite{Matschull1} for open
universes of genus $g=0$ with $n$ massive, but spinless particles.
We adapt the notation and conventions in \cite{Matschull1} to the
conventions in the main text. An important difference in conventions is
the different  sign convention used for the Minkowski metric in
\cite{Matschull1} (``mostly plus'' instead of our ``mostly minus'').
Hence our pairing $\langle J_a,P_b\rangle = \eta_{ab}$ is minus the
trace used in \cite{Matschull1}. Also, there is an overall relative sign
in the definition of the symplectic structure. Finally,
we  use the  group $\PPoi$ instead of the group
$SL(2,\RR)\ltimes \mathfrak{sl}_2$ used in \cite{Matschull1}.

\begin{figure}
\protect\input epsf \protect\epsfxsize=12truecm \protect\centerline{\epsfbox{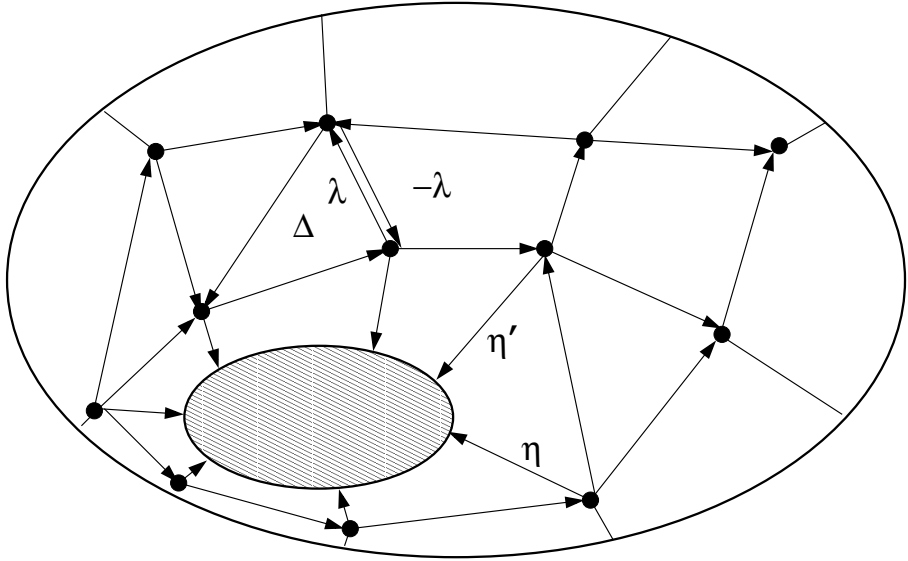}} \caption{ A graph on the spatial surface as used in
\cite{Matschull1}} \label{mgraph3}
\end{figure}

In \cite{Matschull1}, spatial infinity is incorporated as a
connected boundary with an associated boundary condition on
dreibein and spin connection. This boundary condition is the
requirement that the dreibein and spin connection take the form
\eqref{dreib2} associated to a cone with deficit angle $\mu$ and
spin $s$ in a region around the boundary. The formalism relies on
an ADM-decomposition of the spacetime and an oriented graph
$\Gamma$ embedded into a spatial slice. The embedding is such that
every vertex of the graph coincides with one of the particles and
each oriented edge either connects two particles or is a spatial
 half line extending from one particle to the boundary, see
 Fig.~\ref{mgraph3}.
The oriented edges can therefore be grouped into
 two disjoint sets,  the set $\Gamma_+$ of oriented {\em internal}
  edges that start and end at two vertices of the graph and the set
 $\Gamma_\infty$ of {\em external } edges that
 start at a vertex and extend to the boundary. For both sets of edges,
 there are corresponding sets $\Gamma_-$ and $\Gamma_{-\infty}$  which are obtained by reversing the orientation of each edge in $\Gamma_+$ and $\Gamma_\infty$,
 respectively. Among the faces of the graph, one distinguishes {\em internal polygons}, faces for which all sides are edges connecting particles and {\em external polygons}, faces for which at least one side is a component of the boundary.

The symplectic structure is described in terms of a symplectic
potential defined on an extended phase space of link variables
assigned to the oriented graph as follows. To each oriented
internal edge $\lambda\in\Gamma_+$, one associates an element
$(g_\lambda,\bz_\lambda)\in\PPoi$ \bea \label{edgeassoc}
\lambda\in \Gamma_+\mapsto(g_\lambda,\bz_\lambda)\in \PPoi, \eea
and to the corresponding edge $-\lambda$ with reversed orientation
\bea \label{edgeass} -\lambda\in
\Gamma_+\mapsto(g_\lambda^\inv,-\Ad(g_\lambda)\bz_\lambda)\in\PPoi.
\eea Furthermore, each oriented external edge
$\eta\in\Gamma_\infty$ is equipped with four real parameters
$M_\eta, T_\eta, L_\eta$ and $\varphi_\eta$ \bea \label{extedge}
\eta\in\Gamma_\infty\mapsto (M_\eta, T_\eta, L_\eta,
\varphi_\eta). \eea

In terms of these link variables the Lorentz component of the
holonomy around a particle $\pi$ with respect to a base point in
an adjacent face $\Delta$ is given by \bea \label{holform} u_{\pi,
\Delta}=\prod_{\lambda\in\Delta\cap \pi} g_\lambda, \eea where
$\lambda\in \Delta\cap \pi$ denotes the set of all edges that
either start or end at the vertex associated to the particle $\pi$
and are a side of the polygon $\Delta$. The total mass $\mu$ and
total spin $s$ of the universe can be expressed in terms of the
link variables as
 \begin{align} \label{massdef} &\mu=\sum_{\eta\in\Gamma_\infty}
M_\eta\\
 \label{spindef} &s=-\sum_{\lambda\in\Gamma_+} \langle\bz_\lambda-\Ad(g_\lambda) \bz_\lambda, J_0\rangle.
 \end{align}

The symplectic structure on the extended phase space is given
as the exterior derivative of the symplectic potential
\bea
\label{matpot0} -\Theta_{m}(g_\lambda,
\bz_\lambda, M_\eta, T_\eta, L_\eta, \varphi_\eta)=
-\sum_{\eta\in \Gamma_\infty}(T_\eta \delta M_\eta+ L_\eta\delta
\phi_\eta) -\sum_{\lambda\in\Gamma_+}
\langle g_\lambda^\inv \delta g_\lambda ,\bz_\lambda\rangle,
\eea
where we have inserted an overall minus sign in the result given
in  \cite{Matschull1} and taken into account the different  sign convention for
the pairing $\langle, \rangle$
to make  the formula \eqref{matpot0}
 compatible with the conventions used
in this paper.
The physical phase space and its symplectic structure are obtained from the extended phase space by imposing four sets of first-class constraints and by dividing by the associated gauge transformations. The first set of constraints states that for each internal polygon the sum of the relative position vectors $\bz_\lambda$ associated to its sides must vanish
\bea \label{intconst} \mathbf{Z}_\Delta=\sum_{\lambda\in\Delta}\bz_\lambda\approx 0\qquad\forall \Delta\in\Pi_0. \eea
The associated gauge transformations are Lorentz transformations $v\in\LLor$ acting on the edges of the polygon via
$\Delta$
\begin{align}
\label{intgt}
\mathbf{Z}_\Delta:\qquad &g_\lambda\mapsto g_\lambda v^\inv & &\text{for}\;\lambda\in \Delta\\
&g_\lambda\mapsto v g_\lambda & &\text{for}\;-\lambda\in \Delta\nonumber\\
&\bz_\lambda\mapsto \Ad(v)\bz_\lambda & &\text{for}\;\lambda\in\Delta\;\text{or}\;-\lambda\in \Delta\nonumber.
\end{align}
The second set of constraints refers to  external polygons $\Delta$
\bea
\label{extconst}
\mathcal{Z}_\Delta=T_{\eta'}-T_{\eta}+\frac{s}{2\pi}(\varphi_{\eta'}-\varphi_\eta)-\sum_{\lambda\in\Delta\cap
  (\Gamma_+\cup \Gamma_-)}\langle \bz_\lambda, J_0\rangle\approx 0 \qquad\forall \Delta\in\Pi_{\infty},
\eea
where $\eta'$ denotes the edge consecutive to edge $\eta$ in counterclockwise direction, see Fig.~\ref{mgraph3}, and $\Delta\cap
(\Gamma_+\cup \Gamma_-)$ the sides of the external polygon that correspond to internal edges. It generates gauge transformations that shift the parameters $M_\eta$ and $L_\eta$ of the external edges in the polygon $\Delta$ as well as a rotating
all internal polygons by a real parameter $\tau$
\begin{align}
\label{extgt}
\mathcal{Z}_\Delta:\; &M_{\eta'}\mapsto M_{\eta'}-\tau,\;\;M_\eta\mapsto M_\eta+\tau\\
 &L_{\eta'}\mapsto L_{\eta'}+\tfrac{s}{2\pi}\tau,\;\;L_\eta\mapsto L_\eta-\tfrac{s}{2\pi}\tau\nonumber\\
&g_\lambda\mapsto e^{-\frac{\tau}{2\pi}(\varphi_{\eta'}-\varphi_\eta) J_0} g_\lambda e^{\frac{\tau}{2\pi}(\varphi_{\eta'}-\varphi_\eta) J_0},\quad\bz_\lambda\mapsto\Ad( e^{-\frac{\tau}{2\pi}(\varphi_{\eta'}-\varphi_\eta) J_0})\bz_\lambda\;\text{for}\;\lambda,-\lambda\notin\Delta\nonumber\\
&g_\lambda\mapsto e^{-\frac{\tau}{2\pi}(\varphi_{\eta'}-\varphi_\eta) J_0} g_\lambda e^{(\frac{\tau}{2\pi}(\varphi_{\eta'}-\varphi_\eta)-\tau) J_0},\quad\bz_\lambda\mapsto\Ad( e^{(-\frac{\tau}{2\pi}(\varphi_{\eta'}-\varphi_\eta)+\tau) J_0})\bz_\lambda\nonumber\\
&\qquad\qquad\qquad\quad\qquad\quad\qquad\qquad\quad\;\qquad\qquad\qquad\qquad\;\text{for}\;\lambda\in\Delta\cap(\Gamma_+\cup \Gamma_-)\nonumber.
\end{align}
Furthermore, there is an additional constraint for each external
edge $\eta\in\gamma_\infty$, which relates the associated
parameters $M_\eta, L_\eta$ to the total spin of the asymptotic
cone
\bea
\label{angextconst}
\mathcal{J}_\eta=L_\eta
+\tfrac{s}{2\pi} M_\eta\approx 0
\qquad\forall\eta\in\Gamma_\infty.
\eea
The gauge transformations
for these constraints  shift the variables $T_\eta$ and
$\varphi_\eta$ of this edge and,
again, also rotate all internal polygons
\begin{align}
\label{angextgt} \mathcal{J}_\eta:\qquad
&T_\eta\mapsto T_\eta+\tfrac{s}{2\pi}\psi,\quad \varphi_\eta\mapsto
\varphi_\eta
- \psi \\
& g_\lambda\mapsto e^{-\frac{M_\eta}{2\pi}\psi J_0} g_\lambda
e^{\frac{M_\eta}{2\pi}
\psi J_0},\quad\bz_\lambda\mapsto\Ad(
e^{-\frac{M_\eta}{2\pi}\psi J_0})\bz_\lambda\qquad\forall\lambda\in \Gamma_+\cup\Gamma_-\nonumber,
\end{align}
where  $\psi \in\RR$.
Finally, there is set of mass constraints for each particle. Parametrising the holonomy \eqref{holform} around the particle in terms of a momentum three vector  $\bp_{\pi,\Delta}$, these mass constraints read
\bea \label{massconst} C_{\pi,\Delta}= \sqrt{\bp_{\pi,\Delta}^2}-\mu_\pi\approx 0 \qquad\text{where}\quad u_{\pi,\Delta}=e^{-p_{\pi,\Delta}^a
J_a}. \eea
The associated gauge transformations translate the vectors $\bz_\lambda$ of internal edges $\lambda$ starting or
ending at the particle and shift the parameter $T_\eta$ of the external edges starting at the particle
\begin{align}
\label{massgt}
C_{\pi,\Delta}: \qquad &\bz_\lambda\mapsto \bz_\lambda-\tau \hat{\bp}_{\pi,\Delta} & &\text{if}\;\lambda\in \Delta\;\text{and points to vertex}\;\pi\\
&\bz_\lambda\mapsto \bz_\lambda+\tau \hat{\bp}_{\pi,\Delta} & &\text{if}\;\lambda\in \Delta\;\text{and starts at vertex}\;\pi\nonumber\\
& T_\eta\mapsto  T_\eta +\tau \langle \hat{\bp}_{\pi,\Delta}, J_0\rangle & &\text{if}\;\eta\in \Delta\;\text{and starts at
vertex}\;\pi\nonumber,\qquad\tau\in\RR,
\end{align}
with
\bea
\label{phatdef}
 \hat{\bp}_{\pi,\Delta}=\tfrac{1}{\sqrt{\bp_{\pi,\Delta}^2}}\bp_{\pi,\Delta}.
\eea

\subsection{Description for a minimal graph}

\label{minsect}

\begin{figure}
\protect\input epsf \protect\epsfxsize=12truecm \protect\centerline{\epsfbox{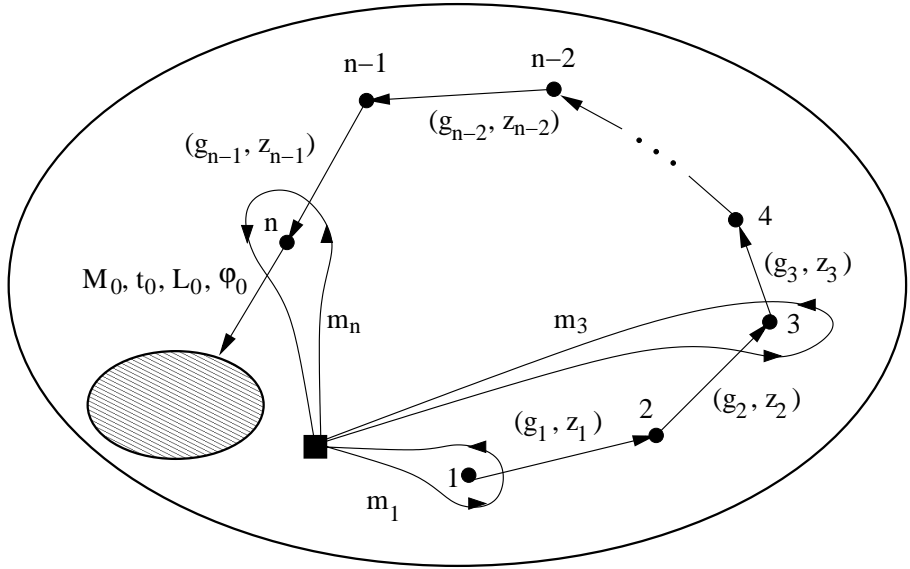}} \caption{ The minimal graph for the surface $S_{0,n}$}
\label{mgraph2}
\end{figure}

We now discuss Matschull's description of the phase space for a minimal graph, which, in a certain sense is the dual of a set of generators of the spatial surface's fundamental group  $\pi_1(S_{0,n}^\infty)$. This minimal graph  consists of $n$ vertices, each coinciding with one of the particles, a single face and $n$ edges, $n-1$ of which connect different particles
and therefore can be viewed as internal, and one external edge, extending from the $n^{th}$ particle to the boundary, as pictured in  Fig.~\ref{mgraph2}.

The phase space variables for this graph are  $n-1$ Poincar\'e elements
$(g_i,\bz_i)\in\PPoi$, $i=1,\ldots,n-1$, associated to each internal edge and the four real parameters  $M_0$, $T_0$, $L_0$, $\varphi_0$ associated to the external edge. In terms of these variables, the Lorentz components $\tilde u_\mi$ of the holonomies along the loops $m_i$ around the particles are given by
\begin{align}
\label{gihols} &\tilde u_\me=g_1^\inv\qquad \tilde u_\mi=g_i^\inv g_{i-1}\qquad i=2,\ldots, n,
\end{align}
where we set $g_n=e^{-\mu J_0}$.
The equation \eqref{massdef} for the mass of the universe reads
 \bea \label{masscond} M_0=\mu, \eea
and equation \eqref{spindef} for its spin
\begin{align}
\label{spingr} s=-\sum_{i=1}^{n-1}\langle (1-\Ad(g_i))\bz_i, J_0\rangle.
\end{align}
The symplectic potential \eqref{matpot0} on the extended phase space becomes
\begin{align}
\label{matpot2} &-\Theta_m(g_i,\bz_i,\mu, s, T_0,\varphi_0 )=-T_0\delta\mu - L_0\delta \varphi_0-\sum_{i=1}^{n-1}\langle g_i^\inv\delta g_i,
\bz_i\rangle.
\end{align}

To obtain the physical phase space, one has to impose the constraints \eqref{intconst}, \eqref{extconst}, \eqref{angextconst} and \eqref{massconst} and to divide by the associated gauge transformations. As the graph in Fig.~\ref{mgraph2} has only a single face and one external edge, the constraints \eqref{intconst} and \eqref{extconst} are
satisfied trivially. However, there is an external edge and therefore an associated constraint of the form \eqref{angextconst}.  Taking into account \eqref{masscond}, this constraint  becomes
\begin{align}
\label{minangextconst} &\mathcal{J}_0=L_0+\frac{\mu s}{2\pi}\approx 0,
\end{align}
where $s$ is the spin of the universe given as a function of the
variables $g_i$, $\bz_i$ by \eqref{spingr}. The associated gauge
transformation \eqref{angextgt} is precisely the transformation
\eqref{spinunivgt} given in the main text.

The remaining constraints are
the $n$ mass constraints \eqref{massconst} the particles.
With the parametrisation $\tilde u_\mi=e^{-
  \tilde p_\mi^a J_a}$ for the Lorentz holonomies \eqref{gihols} around each particle, they read
\bea \label{massconst2} C_{i}:\qquad
\sqrt{\tilde\bp^2_\mi}-\mu_i\approx 0\qquad i=1,\ldots,n, \eea and
the gauge transformations \eqref{massgt} now take the form
 \eqref{minmassgt2} given in the main text.

The physical phase space is obtained from the extended
phase space, parametrised by the $n-1$ Poincar\'e elements $(g_i,\bz_i)$ and the parameters $M_0=\mu$, $T_0$, $L_0$ and $\varphi_0$ and carrying
symplectic potential \eqref{matpot2}, by imposing the $n$ mass constraints \eqref{massconst2} for each particle and the constraint \eqref{minangextconst} and dividing by the associated gauge transformations \eqref{minmassgt2} and \eqref{spinunivgt}.
It therefore has the dimension $6(n-1)+4-2(n+1)=4n-4$.


\end{document}